\documentclass{solarphysics}
\usepackage[optionalrh,solaromanenum]{spr-sola-addons} % For Solar Physics
\usepackage{epsfig}                     % For eps figures, old commands
\usepackage{graphicx}                    % For eps figures, newer & more powerfull
\usepackage{color}                       % For color text: \color command
\usepackage{hyperref}
\usepackage{natbib}
\usepackage{solaheader}

\usepackage{times}
\usepackage{txfonts}
\usepackage{multirow}
\usepackage{rotating}

\begin{document}

\begin{article}

\begin{opening}

\title{A Comparison Between Global Proxies of the Sun's Magnetic Activity Cycle: Inferences from Helioseismology}

\author{A.-M.~\surname{Broomhall}$^{1,2}$ \sep V.M. \surname{Nakariakov}$^{2,3,4}$}

%%%%%%%%%%%%%%%%%%%%%%%%%%%%%%%%%%%%%%%%%%%%%%%%%%%
%% Runningheads
%
\runningauthor{A.-M. Broomhall, V.M. Nakariakov}
\runningtitle{Solar Cycle Variations in Global Proxies}

%%%%%%%%%%%%%%%%%%%%%%%%%%%%%%%%%%%%%%%%%%%%%%%%%%%
%% Affilations
%
  \institute{$^1$Institute of Advanced Studies, University of Warwick, Coventry, CV4 7HS, UK\\
                    email: \url{a-m.broomhall@warwick.ac.uk}\\
                $^2$Centre for Fusion, Space, and Astrophysics, Department of Physics, University of Warwick, Coventry CV4 7AL, UK\\
	     $^3$School of Space Research, Kyung Hee University, Yongin, 446-701, Gyeonggi, Korea\\
                $^4$Central Astronomical Observatory at Pulkovo of RAS, St Petersburg 196140, Russia}

%%%%%%%%%%%%%%%%%%%%%%%%%%%%%%%%%%%%%%%%%%%%%%%%%%%
%%% Abstract
\begin{abstract} The last solar minimum was, by recent standards, unusually deep and long. We are now close to the maximum of the subsequent solar cycle, which is relatively weak. In this article we make comparisons between different global (unresolved) measures of the Sun's magnetic activity, to investigate how they are responding to this weak-activity epoch. We focus on helioseismic data, which are sensitive to conditions, including the characteristics of the magnetic field, in the solar interior. Also considered are measures of the magnetic field in the photosphere (sunspot number and sunspot area), the chromosphere and corona (10.7\,cm radio flux and 530.3\,nm green coronal index), and two measures of the Sun's magnetic activity closer to Earth (the interplanetary magnetic field and the galactic cosmic-ray intensity). Scaled versions of the activity proxies diverge from the helioseismic data around 2000, indicating a change in relationship between the proxies. The degree of divergence varies from proxy to proxy with sunspot area and 10.7\,cm flux showing only small deviations, while sunspot number, coronal index, and the two interplanetary proxies show much larger departures. In Cycle 24 the deviations in the solar proxies and the helioseismic data decrease, raising the possibility that the deviations observed in Cycle 23 are just symptomatic of a 22-year Hale cycle. However, the deviations in the helioseismic data and the interplanetary proxies increase in Cycle 24. Interestingly the divergence in the solar proxies and the helioseismic data are not reflected in the shorter-term variations (often referred to as quasi-biennial oscillations) observed on top of the dominant 11-year solar cycle.  However, despite being highly correlated in Cycle 22, the short-term variations in the interplanetary proxies show very little correlation with the helioseismic data during Cycles 23 and 24.
\end{abstract}

%%%%%%%%%%%%%%%%%%%%%%%%%%%%%%%%%%%%%%%%%%%%%%%%%%%
%% Keywords
%

\end{opening}
%-------------------------------------------------

\section{Introduction}\label{section[introduction]}
The last solar activity cycle (Cycle 23) was unusual in that it was abnormally long ($>12\,\rm years$), 
and the minimum that followed it was noticeably deeper than any observed since the early 1900s. 
Recent work by Livingston, Penn, and co-authors, using infrared measurements, suggest that the 
maximum magnetic-field strength in sunspots has weakened since around 1998 \citep
{2002SoPh..207...41L,2006ApJ...649L..45P,2011IAUS..273..126P,2012ApJ...757L...8L,2014ApJ...787...22W}. These results, however, remain controversial, with other studies showing that 
the most dominant observed trends are simply those associated with the regular solar-cycle variation 
\citep{2011A&A...533A..14W,2011ApJ...742L..36P,2012A&A...541A..60R,2014SoPh..289..593P}. 
\inlinecite{2012ApJ...758L..20N} found that a long-term decline in the sunspot field strength was 
observed when both large and small sunspots were included in the data but that only solar-cycle 
variations were present when only sunspots with the strongest magnetic fields were included. 
\inlinecite{2012ApJ...758L..20N} explain this in terms of a change in the fraction of large and 
small sunspots. 

Inspired by this controversy, comparisons have been made between sunspot number (SSN) variation, 
which is a measure of the number of spots on the visible solar disk, and other global proxies of the 
Sun’s magnetic field, such as the 10.7cm radio flux (F$_{10.7}$). Many of these comparisons show that 
proxies that were in good agreement with each other in previous solar cycles (before Solar Cycle 23) 
are now diverging \citep{2012ApJ...757L...8L,2012JSWSC...2A..06C}. It is now well-known that biases 
exist in the sunspot-number statistics (see \opencite{2014SSRv..186...35C} for a recent review). 
Although corrections to the sunspot-number trend do reduce some of the observed deviations between 
the SSN and F$_{10.7}$, significant departures persist after 2002, implying that the discrepancy 
indicates a true change on the Sun \citep{2011A&A...536L..11L,2014SSRv..186...35C}. 

\inlinecite{2014SSRv..186...35C} also observed a decline in the ratio of the observed SSN and a 
synthetic SSN constructed from the Mount Wilson Observatory Magnetic Plage Strength Index: Solar-
cycle variations were visible in the ratio but there was also an underlying systematic decline over 
the past two solar cycles. Furthermore, \inlinecite{2014SSRv..186...35C} also demonstrated a decrease 
in the average number of spots per sunspot group in Cycles 23 and 24, which is consistent with the 
results of \inlinecite{2013Ge&Ae..53..953T}. \inlinecite{2011SoPh..270..463J} observed a significant 
change in the growth and decay rates of sunspot groups since the beginning of Cycle 23. These 
results are consistent with a number of results that demonstrate a deficit in the number of small 
spots observed at the maximum of Cycle 23 compared to Cycle 22 \citep{2011ApJ...731...30K,2011A&A...536L..11L,2012JSWSC...2A..06C}, since small sunspots are known to be relatively short-lived. However, we note that  \inlinecite{2012ApJ...758L..20N} observed an increase in the number of 
small sunspots and a decrease in the number of large sunspots since around 2006. 

The solar surface and overlying atmosphere are not the only places where a change in behaviour has 
been observed. The solar interior also appears to be behaving differently in recent years given 
expectations based on previous observations. Helioseismology allows the interior of the Sun to be 
probed by observing the properties of the Sun's natural acoustic oscillations ($p$-modes). Each 
oscillation is trapped within a well-defined region of the solar interior, known as a cavity. Solar 
$p$-modes are sensitive to conditions in these cavities, and as such are affected by the presence of 
flows and magnetic fields (\textit{e.g.} \opencite{1985Natur.318..449W}; \opencite{1989A&A...224..253P}; 
\opencite{1990Natur.345..322E}). One such flow is the meridional circulation, which transports 
near-surface material poleward and is a key component of many solar-dynamo theories. The meridional -flow speed measured in the last solar minimum (between Cycles 23 and 24) was larger than the flow 
speed observed in the minimum between Cycles 22 and 23 \citep{2010ApJ...717..488B, 
2010Sci...327.1350H}. Another well-known flow associated with the solar interior is the torsional 
oscillation, which is also behaving differently than in the same phase of the previous cycle: The 
poleward branch of the torsional oscillation that will eventually be associated with Cycle 25 is 
weaker than expected, given the observations of the previous cycle \citep{2013ApJ...767L..20H}.  
\inlinecite{2012ApJ...758...43B} demonstrated that the low-frequency low-degree [$\ell$] $p$-modes showed 
almost no solar-cycle associated change in frequency in Cycle 23, despite showing a noticeable 
variation in Cycle 22. \citeauthor{2012ApJ...758...43B} explained this in terms of a thinning of the 
near-surface ``magnetic layer'' in Cycle 23, which is believed to be responsible for the observed 
perturbations in frequency with solar cycle. More recently, \inlinecite{2015arXiv150207607S} demonstrated that the frequencies of high-frequency modes varied by 30\% less in the rising phase of Cycle 24 than in Cycle 23, which is in agreement with surface and atmospheric measures of the Sun's magnetic field. However, the frequencies of the low-frequency modes appear to change by the same amount in Cycle 23 and the rising phase of Cycle 24, implying that below 1400\,km the magnetic field has not changed.

\inlinecite{2012ApJ...758...43B} also showed that at higher frequencies $(\ge 2450\,\rm\mu Hz)$ the 
frequencies of the low-$\ell$ $p$-modes behaved in a similar manner in Cycles 22, 23, and the rising 
phase of Cycle 24. As such, the frequencies of these oscillations can be considered as proxies of 
the Sun's magnetic-activity cycle. We define the helioseismic frequency shift as the mean change in 
frequency with time. The method by which this frequency shift is determined will be described in 
Section \ref{section[heliodata]}. Such a helioseismic proxy would be sensitive to the effect of a 
magnetic field on the solar interior, although it is primarily the influence of the magnetic field 
in near-surface regions that affects the $p$-mode oscillations (\textit{e.g.} \opencite{1990Natur.345..779L}). 
If we consider the helioseismic frequency shifts to be a global proxy of the Sun's magnetic field, it 
is only natural to compare the helioseismic frequency shifts to other commonly used global 
proxies of the Sun's magnetic activity. The observed frequency shifts can be explained in terms of 
direct (the Lorentz force) and indirect effects (changes in the properties of the cavities). To date 
there is no agreement as to the exact balance of these two effects (\textit{e.g.} \opencite
{1986Natur.323..603R}; \opencite{1994SoPh..152..261J}; \opencite{2005ApJ...625..548D}). Comparisons 
with other activity proxies may help to constrain the nature of the observed frequency shifts.

Comparisons between the activity proxies have been performed previously  (\textit{e.g.} 
\opencite{1998A&A...329.1119J};  \opencite{2001MNRAS.327.1029A}; \opencite{2001SoPh..200....3T}; 
\opencite{2002ApJ...580.1172H}; \opencite{2004MNRAS.352.1102C}; \opencite{2007ApJ...659.1749C}; 
\opencite{2009ApJ...695.1567J}; \opencite{2012A&A...545A..73J}). \inlinecite{2007ApJ...659.1749C} 
compared six different proxies with the Sun-as-a-star frequency shifts and demonstrated that better 
correlations are observed with proxies that measure both the strong and the weak components of the 
Sun's magnetic field, such as the 10.7 cm radio flux (as opposed the the sunspot number, which 
mainly samples the strong magnetic flux). The weak component of the solar magnetic flux is 
distributed over a wider range of latitudes than the strong component and so a possible explanation 
of these results is in terms of the latitudinal distribution of the modes used in this study (i.e. 
low-degree modes).  \inlinecite{2002ApJ...580.1172H} showed that it is possible to use latitudinal 
inversion techniques to localize the frequency shifts in latitude and, in fact, reconstruct the 
evolution of the surface magnetic field. However, this requires spatially resolved data, such as those 
obtained by the Global Oscillation Network Group (GONG), which has only been collecting data since 
1995. The extended solar minimum between Cycles 23 and 24 means that, to date, this only covers approximately one-and-a-half activity cycles, and so cycle-to-cycle comparisons are not yet possible.

We will compare the relationships between the proxies and the helioseismic frequencies 
observed during the entirety of Cycles 22 and 23, and in the rising phase of Cycle 24. Like 
\inlinecite{2007ApJ...659.1749C}, we compare the helioseismic frequencies to the sunspot number and 
10.7\,cm flux. However, we also consider sunspot-area data and the coronal index. Furthermore, we 
consider two interplanetary proxies of magnetic field (the interplanetary magnetic field and the 
galactic cosmic-ray flux). We use Birmingham Solar Oscillations Network (BiSON: \opencite{1996SoPh..168....1C}) data, which has been 
concatenated using new, improved procedures that are described by \citet{2014MNRAS.441.3009D}. We will perform a comparison 
between the quasi-biennial oscillation (QBO), which is known to exist in the helioseismic 
frequencies \citep{2009ApJ...700L.162B}, and the solar and interplanetary proxies \citep{2014SSRv..186..359B}. A dedicated comparison between the features of the QBO observed in 
helioseismic data and those observed in such a wide range of activity proxies has never before been 
performed.

The remainder of this article is structured as follows: Section \ref{section[data]} describes the data 
used in the article. We describe the helioseismic data and the method by which the frequency shifts 
are obtained (Section \ref{section[heliodata]}). We also describe the other global proxies 
considered in this article (Section \ref{section[proxydata]}). Section \ref{section[comparison]} 
describes the comparison of the global proxies with the helioseismic data. We initially consider the 
differences observed from one 11-year solar cycle to the next. However, we go on to consider 
shorter-term variations in the solar magnetic field, and the consistency of these variations across 
the different proxies (Section \ref{section[2yr]}). Finally, we discuss the implications of these 
results in Section \ref{section[discussion]}.

\section{Data}\label{section[data]}

\subsection{Helioseismic Data}\label{section[heliodata]}

The Birmingham Solar Oscillations Network (BiSON, \opencite{1996SoPh..168....1C}; \opencite{2014MNRAS.441.3009D}) consists of six sites that make line-of-sight velocity observations of the Sun. BiSON makes unresolved, Sun-as-a-star 
observations and so are only sensitive to low-$\ell$ $p$-modes. However, these are the oscillations that 
sample deep layers of the solar interior.  The 
earliest data observed by individual sites are sporadic and so will not be used here. Nevertheless, 
BiSON has now been observing the Sun continuously for over 30\,years, making it the longest 
helioseismic data set (\href{}{\textsf{bison.ph.bham.ac.uk/}}) available and consequently ideal for 
studies of more than one solar cycle. The period 1 January 1985 to 26 March 2014 
will be considered, which covers approximately two-and-a-half solar cycles. 

When using helioseismic data to study the solar cycle, a balance must be made concerning the length 
of time series considered. One must use time series that provide enough resolution in the frequency 
domain to accurately and precisely obtain the frequencies of the oscillations. However, one must 
also use time series short enough to be able to study the evolution of helioseismic parameters in 
time. In order to study variations associated with the 11-year cycle we have used 114 consecutive 
time series of 365-days length that are separated by 91.25\,days. However, when considering shorter-term 
variations we used shorter time series, of length 182.5\,days (but still separated by 91.25\,days). 
Although the precision with which we could obtain the frequencies of the oscillations was reduced by 
this shortening of the time series, this gives a better resolution with which to study variations on 
timescales less than 11\,years. The extracted parameters are essentially the average values observed 
over the 365-day (or 182.5-day) time period under consideration. 

The frequencies of each $p$-mode $[\nu_{l,n}]$ were obtained by fitting an asymmetric profile \citep{1998ApJ...505L..51N} to the frequency power spectra using a maximum-likelihood-estimation technique 
\citep{2009ApJ...694..144F}. Only $p$-modes in the range $2400\le\nu\le3500\,\rm\mu Hz$ were considered 
here for the following reasons: Firstly, these are the most prominent oscillations meaning their 
frequencies can be obtained accurately. Secondly, as we move to higher frequencies the lifetimes of 
the oscillations decrease, meaning they have a wider profile in the power spectrum from which the 
frequencies are obtained. As a result, the precision with which we can obtain the frequencies of the 
oscillations decreases with increasing frequency. Finally, this range includes eight overtones of 
each degree under consideration i.e. there are eight $\ell=0$ modes, eight $\ell=1$ modes, and eight $\ell=2$ modes in 
this frequency range. Only modes with $0\le \ell\le 2$ were considered: The BiSON data have no spatial 
resolution and so the amplitudes of the modes in the power spectrum with $l\ge3$ are low, making it 
hard to determine the frequencies of modes accurately and precisely.We note that the early BiSON data (before approximately 1993) have a relatively low duty cycle as the full BiSON network was not established. Furthermore, because of improvements to the instruments over time, the more recent data are of better quality. This is reflected in the uncertainties associated with the frequencies. In order to compensate for the low duty cycle,  when obtaining the mode frequencies, the window function of the data was convolved with the model fitted to the power spectrum \citep{2009ApJ...694..144F}. As we only consider the most prominent modes in the frequency spectrum they stll have significant signal-to-noise in the early data. Nevertheless, each fitted function was checked by eye to ensure the frequencies obtained were robust.

Once the frequencies were obtained for each of the 365\,day (or 182.5\,day) time series an average set 
of reference frequencies $[\overline{\nu_{n,l}}]$ was determined. This reference set gives the 
weighted average frequencies observed across the entire 30-year epoch under consideration, where the 
weights were determined by the uncertainties given by the maximum likelihood fit to the power 
spectrum. We note that, while the exact choice of time series from which the reference set is constructed can affect the absolute value of frequency shift, the conclusions of this article are not dependent on it. Here, the reference set includes time series obtained when the duty cycle of the data was low and the quality of the observations poorer. However, excluding these from the reference set by, for example, only considering data observed during Cycles 23 and 24, alters the subsequently determined frequency shift by less than $0.5\sigma$. Furthermore, this change is partially explained by a systematic offset caused simply by considering a different time span, and, therefore, a different range of magnetic-activity levels. In real terms, the influence of including the early data is, therefore, negligible.

The frequency shift of an individual mode, in any single 365\,day (or 182.5\,day) timeseries
$[\delta\nu_{n,l,i}$, where $i$ denotes the timeseries under consideration] is then determined by 
subtracting the relevant reference frequency from the frequency observed in that individual time 
series, i.e.
\begin{equation}\label{equation[shift]}
    \delta\nu_{n,\ell,i}=\nu_{n,\ell,i}-\overline{\nu_{n,\ell}}.
\end{equation}
We then determine the weighted-average frequency shift $[\overline{\delta\nu_i}]$ for all modes with 
$0\le \ell\le 2$ and  $2400\le\nu\le3500\,\rm\mu Hz$.  Each frequency shift has an associated 
uncertainty $[\sigma_i]$. We compared these frequency shifts to other well-known proxies of the solar 
magnetic field, which we now describe.

\subsection{Global Proxy Data}\label{section[proxydata]}

Perhaps the best-known global-activity proxy is the sunspot number (SSN), which is a measure of the 
number of sunspot groups and individual sunspots on the solar surface. Here we use the NOAA 
International Sunspot Number (\href{http://www.ngdc.noaa.gov/stp/space-weather/solar-data/solar-
indices/sunspot-numbers/international/}{\textsf{www.ngdc.noaa.gov}}: see discussion by \opencite{2014SSRv..186...35C}). Although 
historically important because of its longevity, the SSN has occasionally been criticised for its 
lack of physicality. Therefore we also consider sunspot area (\href{http://solarscience.msfc.nasa.gov/greenwch.shtml}{\textsf{solarscience.msfc.nasa.gov/greenwch.shtml}}: SA: \opencite
{2010LRSP....7....1H}), as an alternative measure of the photospheric magnetic flux. The 10.7cm 
radio flux (\href{www.ngdc.noaa.gov/stp/space-weather/solar-data/solar-features/solar-
radio/noontime-flux/penticton/}{\textsf{www.ngdc.noaa.gov}}: F$_{10.7}$) is a proxy for the solar activity in the upper chromosphere 
and lower corona, and is sensitive to both strong and weak magnetic-field regions \citep{1987JGR....92..829T}. F$_{10.7}$ is expressed in Radio Flux Units [RFU: $\rm1\,RFU=10^{-22}\,W\,m^{-
2}\,Hz^{-1}$]. The coronal index (\href{http://www.ngdc.noaa.gov/stp/space-weather/solar-
data/solar-indices/solar\_corona/coronal-index/}{\textsf{www.ngdc.noaa.gov}}: CI: \opencite{2005AdSpR..35..410M}) 
is a measure of the total power emitted by the green corona at 530.3\,nm, which corresponds to the 
Fe {\sc xiv} spectral line. The above proxies are observe at different levels in the solar atmosphere 
because they are sensitive to different temperatures, and therefore, indirectly, the magnetic-field 
strength. Consequently, they can be considered as proxies of the solar magnetic field in some region 
of the solar atmosphere. In the remainder of the article we refer to these four proxies as solar 
proxies. We also compare the helioseismic data with two more remote measures of the Sun's magnetic 
field, which we refer to as interplanetary proxies. We consider the interplanetary magnetic field 
(IMF) as extracted from NASA/GSFC's OMNI data set through OMNIWeb
(\href{http://omniweb.gsfc.nasa.gov/html/ow\_data.html}{\textsf{omniweb.gsfc.nasa.gov}}: \opencite{2005JGRA..110.2104K}). We also 
consider the galactic cosmic-ray intensity (CR) obtained from the Moscow neutron monitor
(\href{http://cosrays.izmiran.ru/main.htm}{\textsf{cosrays.izmiran.ru/main.htm}}: \opencite{2013CosRe..51...29B}). Averages of all of 
the proxies were taken over the same 365-day (or 182.5-day) time periods from which the helioseismic 
frequencies were determined. 

\section{Comparison of Global Proxies of the Sun's Magnetic Activity Cycle with Helioseismic 
Frequency Variations}\label{section[comparison]}

\begin{figure}
  \centering
  \includegraphics[clip, width=0.45\textwidth]
{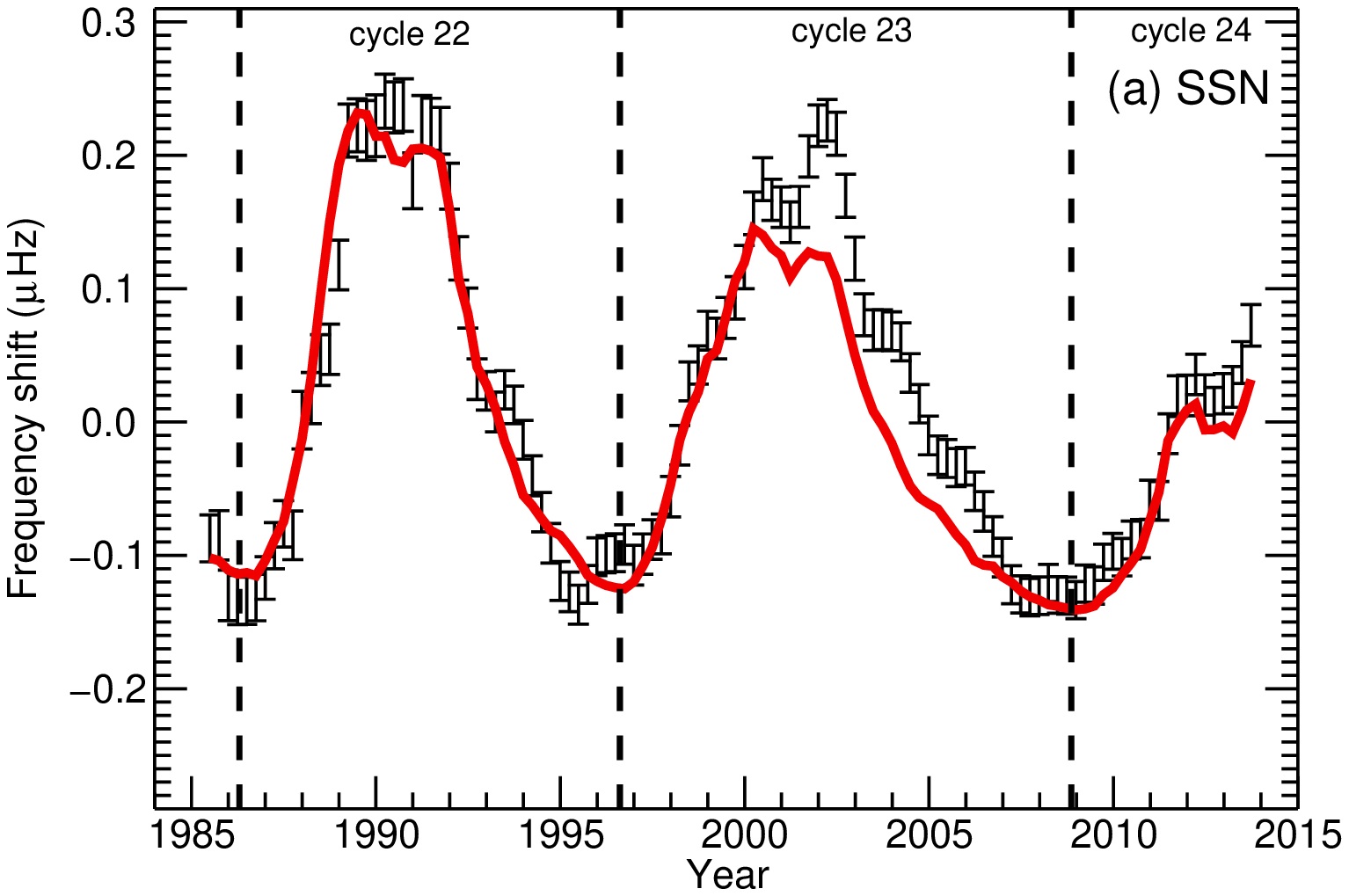}\hspace{1cm}
  \includegraphics[clip, width=0.45\textwidth]{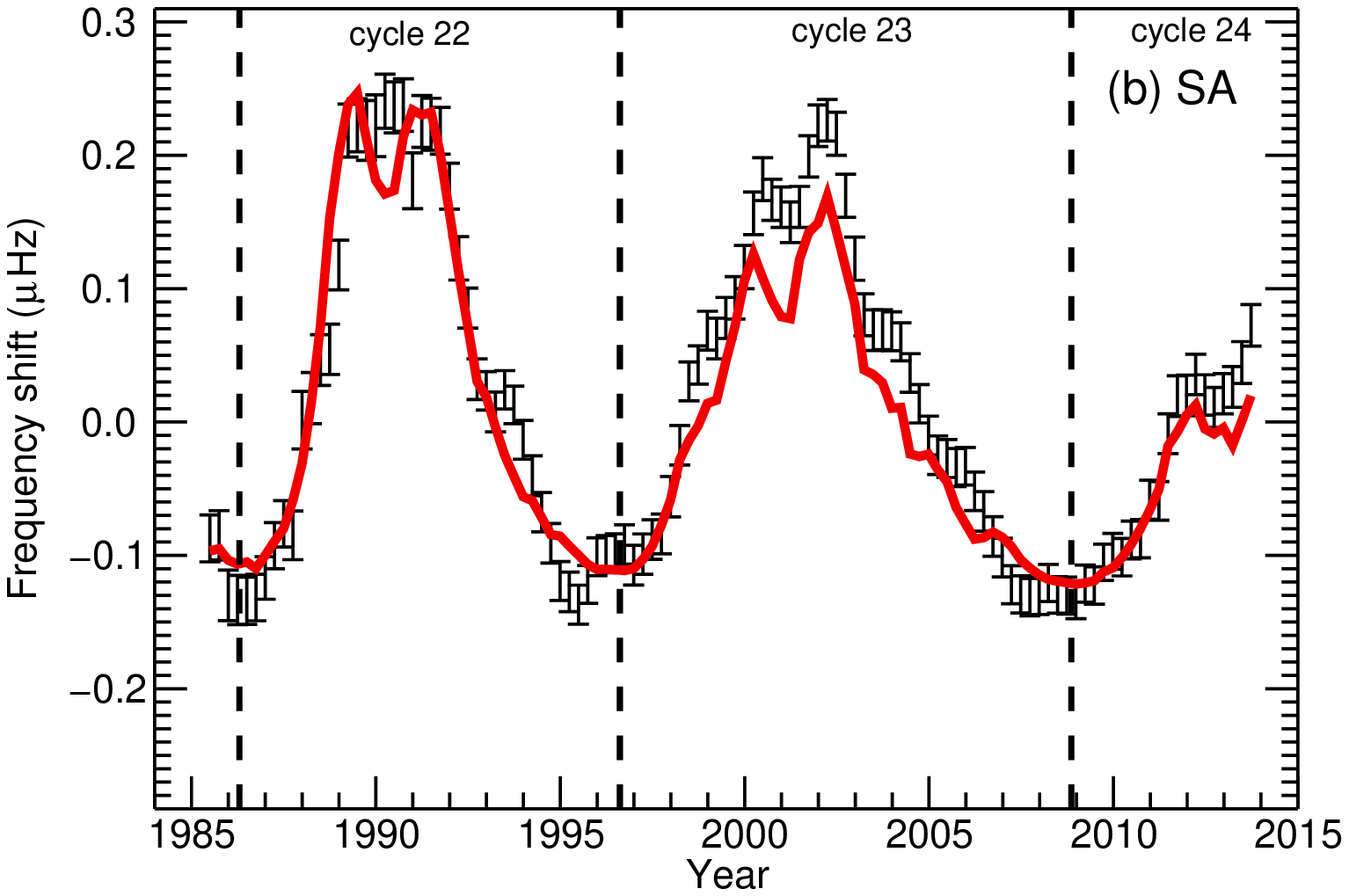}\\
  \includegraphics[clip, width=0.45\textwidth]
{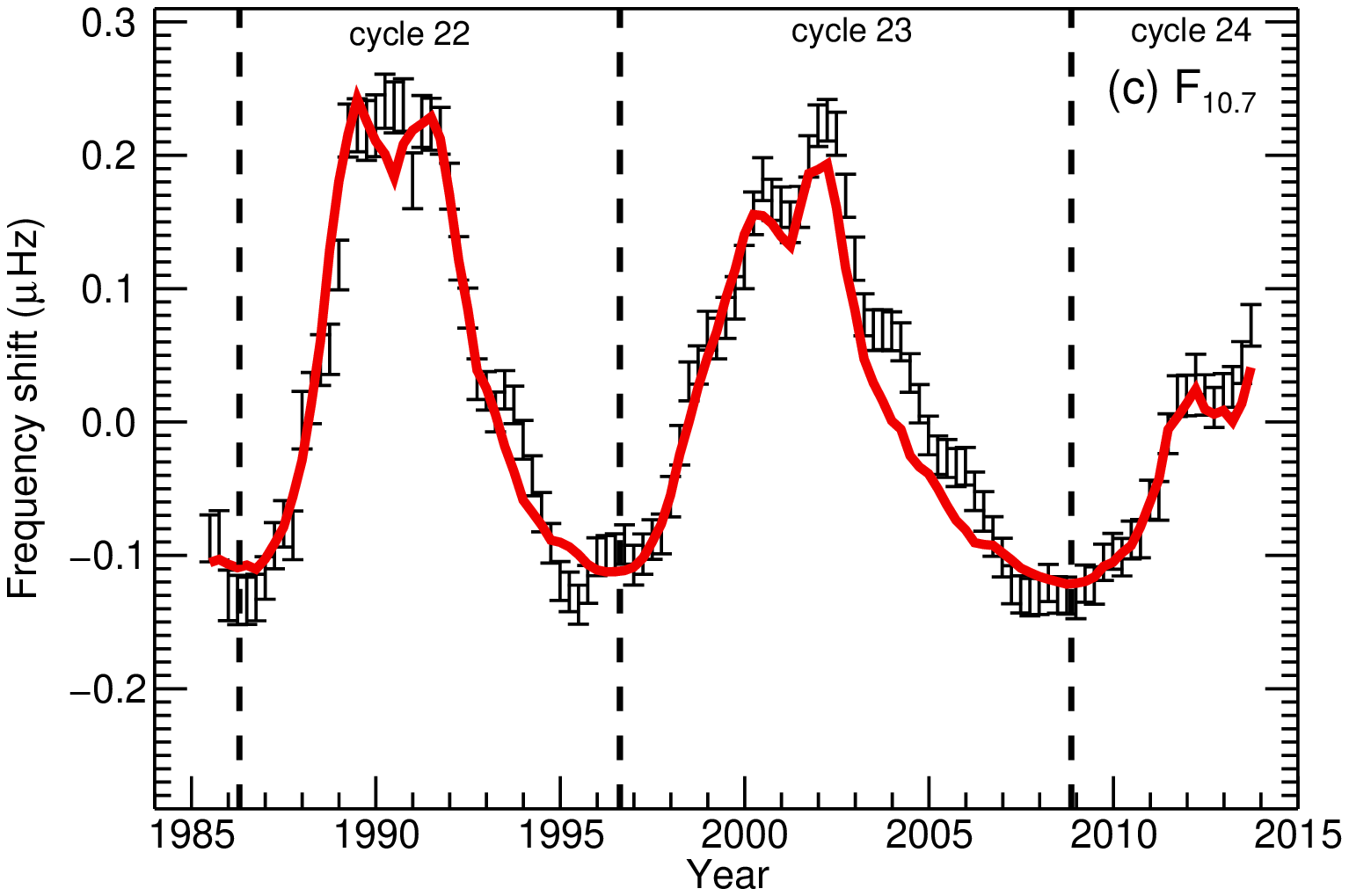}\hspace{1cm}
  \includegraphics[clip, width=0.45\textwidth]{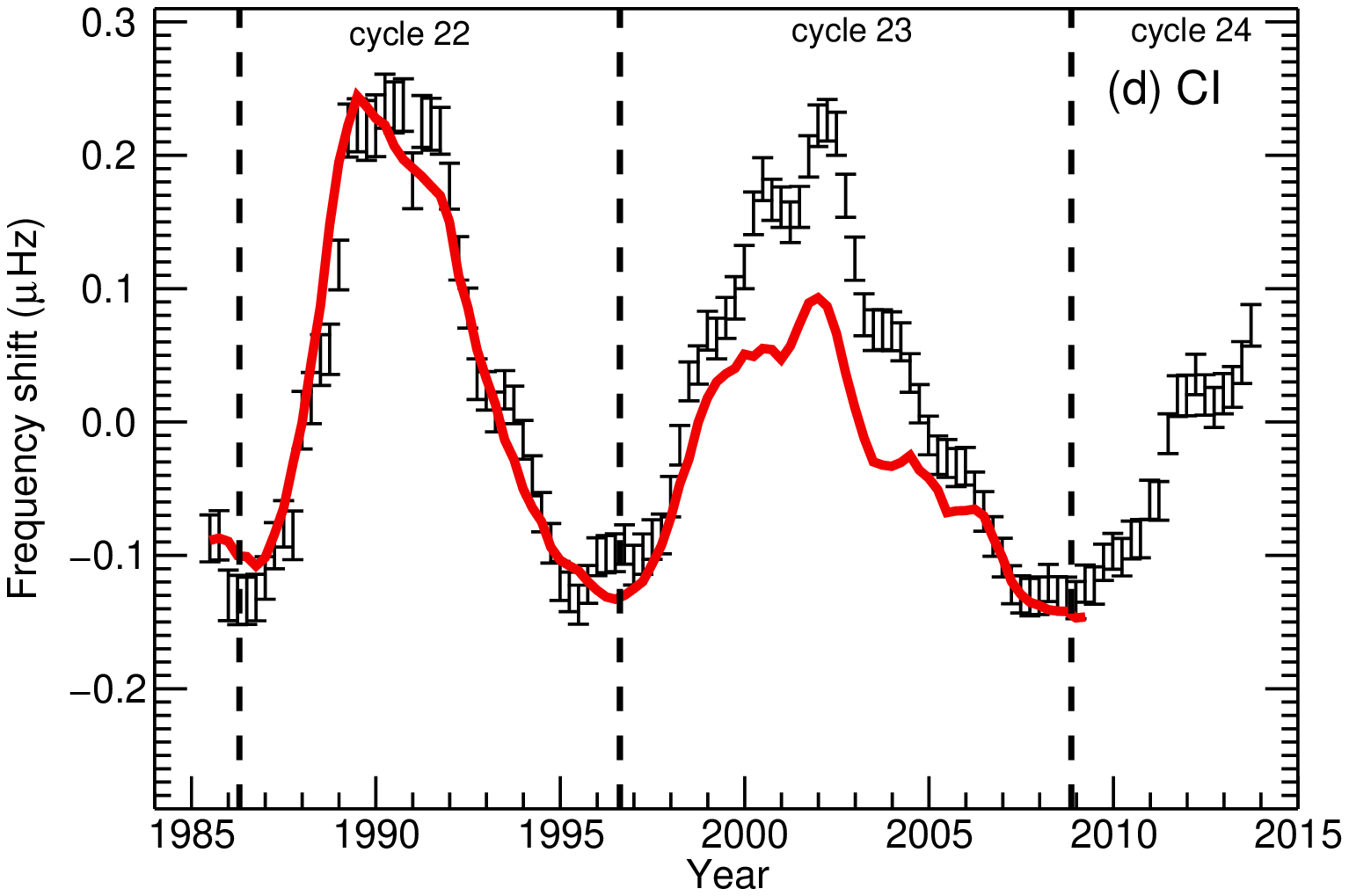}\\
  \includegraphics[clip, width=0.45\textwidth]
{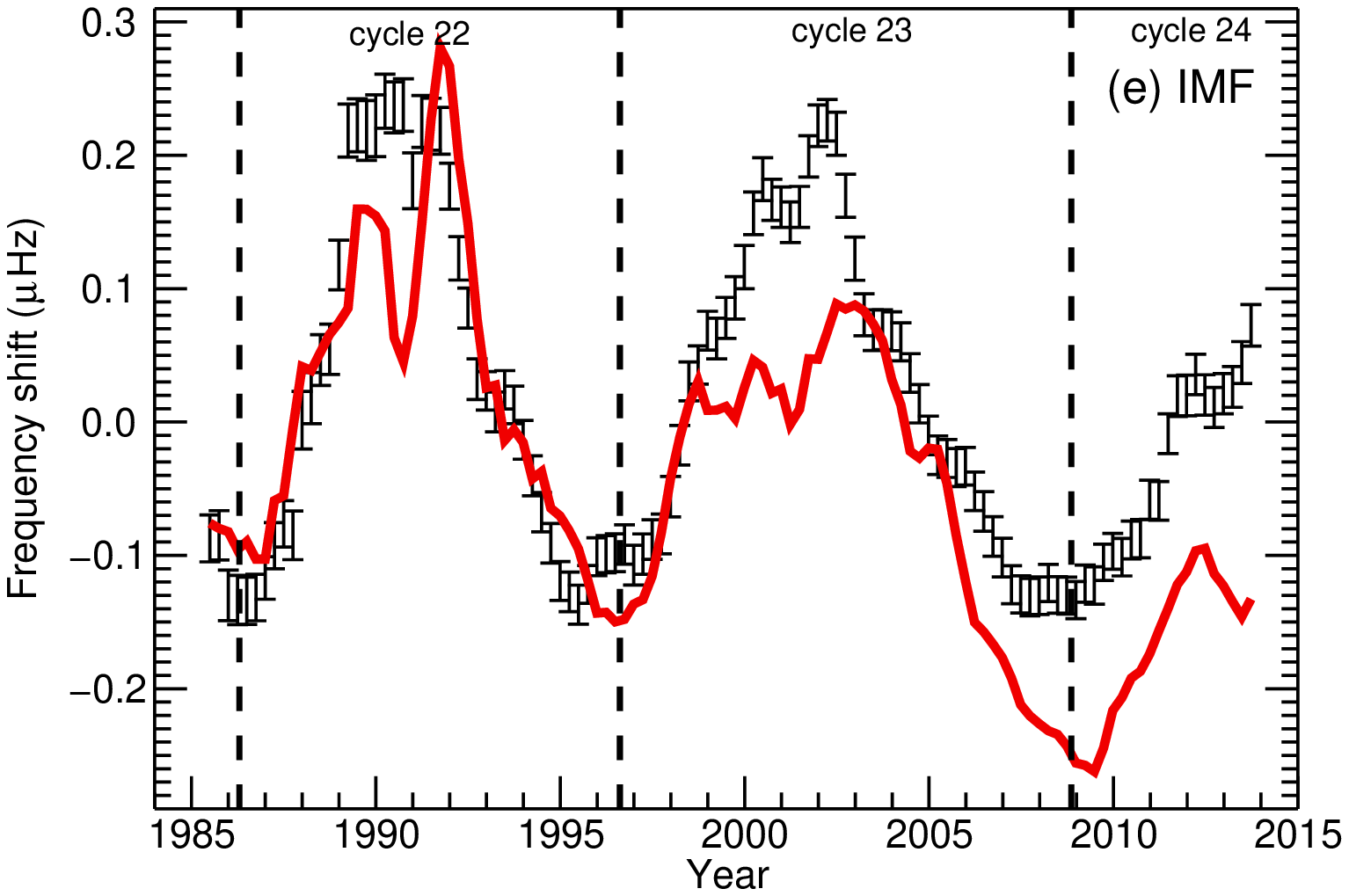}\hspace{1cm}
  \includegraphics[clip, width=0.45\textwidth]{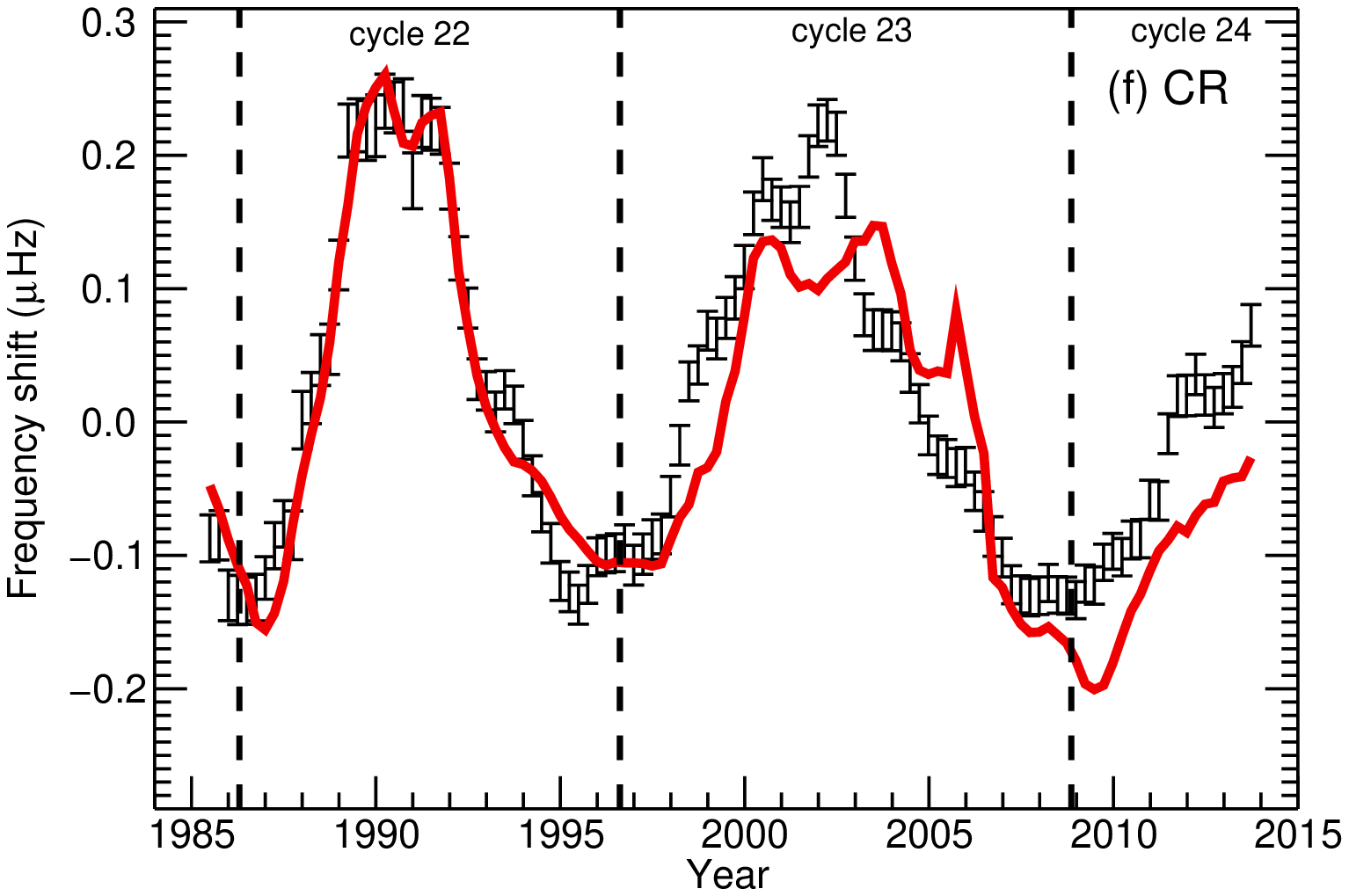}\\
  \caption{Averaged frequency shifts of helioseismic modes (data with plotted uncertainties) obtained using $0\le\ell\le 2$  and $2400\le\nu\le 3500\,\rm\mu Hz$. The results were obtained from 365-day BiSON time 
series, which are separated by 91.25\,days. Red, solid lines represent linearly regressed versions of other 
activity proxies and the coefficients of the linear regressions can be found 
in Table \ref{table[fits]}. Since CR varies in antiphase with solar cycle the 
gradient of the linear regression for CR (panel f) is negative. Panel a: Scaled sunspot number. Panel b: Scaled sunspot area. 
Panel c: Scaled 10.7\,cm radio flux. Panel d: Scaled coronal index. Panel e: Scaled interplanetary magnetic field. 
Panel f: Inversely scaled galactic cosmic ray intensity. The vertical dashed lines separate different solar cycles.}
  \label{figure[shifts_time]}
\end{figure}

Figure \ref{figure[shifts_time]} plots the average $p$-mode frequency shifts $[\overline{\delta\nu_i}]$ 
as a function of time. The amplitudes of the minimum-to-maximum variation of the proxies (including the helioseismic frequency shifts) are very 
different. For example, an $\ell=0$ mode 
with a frequency of $3000\,\rm\mu Hz$ experiences a shift in frequency of approximately $0.4\,\rm\mu 
Hz$ between solar minimum and maximum, whereas the sunspot number changes by over 100 and the 
sunspot area increases by of the order of 2000 millionths of a solar hemisphere (msh). Furthermore, 
CR varies in antiphase with solar cycle, such that the intensity is at a maximum at solar minimum. To 
allow a comparison of the frequency shifts with the proxies a linear regression between the global 
proxies and the frequency shifts was performed. Only values observed during Cycle 22 (between 
1 January 1985 and 29 December 1996) were used in the regression.  This time span includes data from when the duty cycle was relatively low and thus the quality relatively poor in comparison to more recent observations. However, the aim of this paper is to determine whether the relationship between proxy and helioseismic frequency shift varies from one cycle to the next. In this respect the conclusions of this article would not be altered by scaling with respect to a different cycle: One could scale with respect to Cycle 23 and determine the degree of divergence observed in Cycle 22.

The coefficients for the linear regressions can 
be found in Table \ref{table[fits]}. Since the data are overlapping, Monte Carlo simulations were used to determine the uncertainties associated with the linear regressions. These simulations added randomly-generated, Gaussian-distributed noise to the frequency shifts, where the standard deviation of the random numbers were based on the associated uncertainties. The randomly-generated noise accounted for the fact that the uncertainties of adjacent points ($i$ and $i+1$ say) have a 75\% correlation and that this decreases to 50\% at $i+2$, 25\% at $i+3$, and is zero at $i+4$. This linear 
regression has been used to scale the global proxies and these scaled proxies are also plotted as a 
function of time in Figure \ref{figure[shifts_time]}. Note that CR is inversely scaled, and that the inversely scaled CR varies in phase with the other proxies. The scaled proxy values $[\delta\nu_{\textrm{\scriptsize{p}}}]$
observed during Cycles 23 and 24 can be regarded as retrospective predictions of the frequency 
shifts we would expect to observe given the relationship between the proxies and the frequency 
shifts observed during Cycle 22. 

\begin{table*}\caption{Coefficients of linear regressions between helioseismic-frequency shifts and 
global proxies for the two complete solar cycles under consideration. The units in which the 
gradient of the linear regression is measured depend on the proxy under consideration. We have used 
proxy unit (PU) as a global proxy unit where PU is equivalent to msh for SA; RFU for F$_{10.7}$; $\rm W
\,sr^{-1}$ for CI; nT for IMF; impacts min$^{-1}$ for CR.We recall that SSN is a number with no units.  
}\label{table[fits]}
\begin{tabular}{ccccccc}
  \hline
  Proxy & Cycle & Gradient & Intercept & Reduced & Gradient & Intercept \\
   & & $\times10^{-3}$ & & $\chi^2$ & Difference & Difference \\
   & & $[\mu\rm Hz\,PU^{-1}]$ & $[\rm\mu Hz]$ & & $[\sigma]$ &  $[\sigma]$ \\
  \hline
   \multirow{2}{*}{SSN} & 22 & $2.39\pm0.10$ & $-0.145\pm0.008$ & 3.16 &\multirow{2}{*}{$2.0$} 
&  \multirow{2}{*}{$2.0$} \\
   & 23 & $2.67\pm0.10$ & $-0.124\pm0.007$ & 4.74 & & \\
  \hline
  \multirow{2}{*}{SA} & 22 & $0.144\pm0.006$ & $-0.123\pm0.007$ & 3.83 & \multirow{2}{*}{$4.0$} & \multirow{2}{*}{$0.2$} \\
   & 23 & $0.181\pm0.007$ & $-0.125\pm0.006$ & 1.80 &  & \\
  \hline
   \multirow{2}{*}{F10.7} & 22 & $2.51\pm0.10$ & $-0.293\pm0.013$ & 2.73 & \multirow{2}{*}
{$0.9$} & \multirow{2}{*}{$0.1$} \\
   & 23 & $2.64\pm0.10$ & $-0.291\pm0.012$ & 2.89 & & \\
  \hline
   \multirow{2}{*}{CI} & 22 & $24.3\pm1.0$ & $-0.181\pm0.009$ & 3.55 & \multirow{2}{*}{$7.4$} 
& \multirow{2}{*}{$1.5$} \\
   & 23 & $36.4\pm1.3$ & $-0.200\pm0.009$ & 3.01 & \\
  \hline
   \multirow{2}{*}{IMF} & 22 & $91.7\pm3.9$ & $-0.620\pm0.022$ & 12.69 & \multirow{2}{*}{$0.7$} & \multirow{2}{*}{$2.9$} \\
   & 23 & $88.2\pm3.6$ & $-0.529\pm0.022$ & 12.03 & & \\
  \hline
   \multirow{2}{*}{CR} & 22 & $-0.239\pm0.010$ & $2.11\pm0.08$ & 3.00 & \multirow{2}{*}{$1.7$} 
& \multirow{2}{*}{$1.6$} \\
   & 23 & $-0.216\pm0.009$ & $1.93\pm0.08$ & 14.76 & & \\
  \hline
\end{tabular}
\end{table*}

We first consider the SSN.  Up until around 2000, the agreement is good but in the declining phase of 
Cycle 23 discrepancies emerge, although the agreement does appear to improve again in Cycle 24. The 
helioseismic frequency shifts are more compatible with the sunspot area and F$_{10.7}$ predictions, both 
of which show reasonably good agreement with the helioseismic frequencies throughout. As we move 
farther out in the Sun's atmosphere by considering the coronal index we can see that the predicted 
values differ substantially from the observed frequency shifts during Cycle 23: The agreement is 
poor for the majority of Cycle 23. Unfortunately measurements of the coronal index ceased in 2008 
and so we cannot make this comparison in Cycle 24. However, this highlights the need to maintain 
long-duration observations of the Sun at multiple wavelengths. Similarly large departures between 
the observed frequency shifts and the frequency shifts predicted by the IMF are observed in Cycles 
23 and 24. However, we note that, for the IMF, the agreement between the observed and predicted 
frequency shifts during Cycle 22 is poor in comparison with the other proxies. Although the 
helioseismic-frequency shifts and the cosmic-ray predictions (based on the inverse scaling) are more consistent than the IMF, 
significant departures remain in both Cycles 23 and 24.

\begin{figure}
  \centering
  \includegraphics[clip, width=0.45\textwidth]
{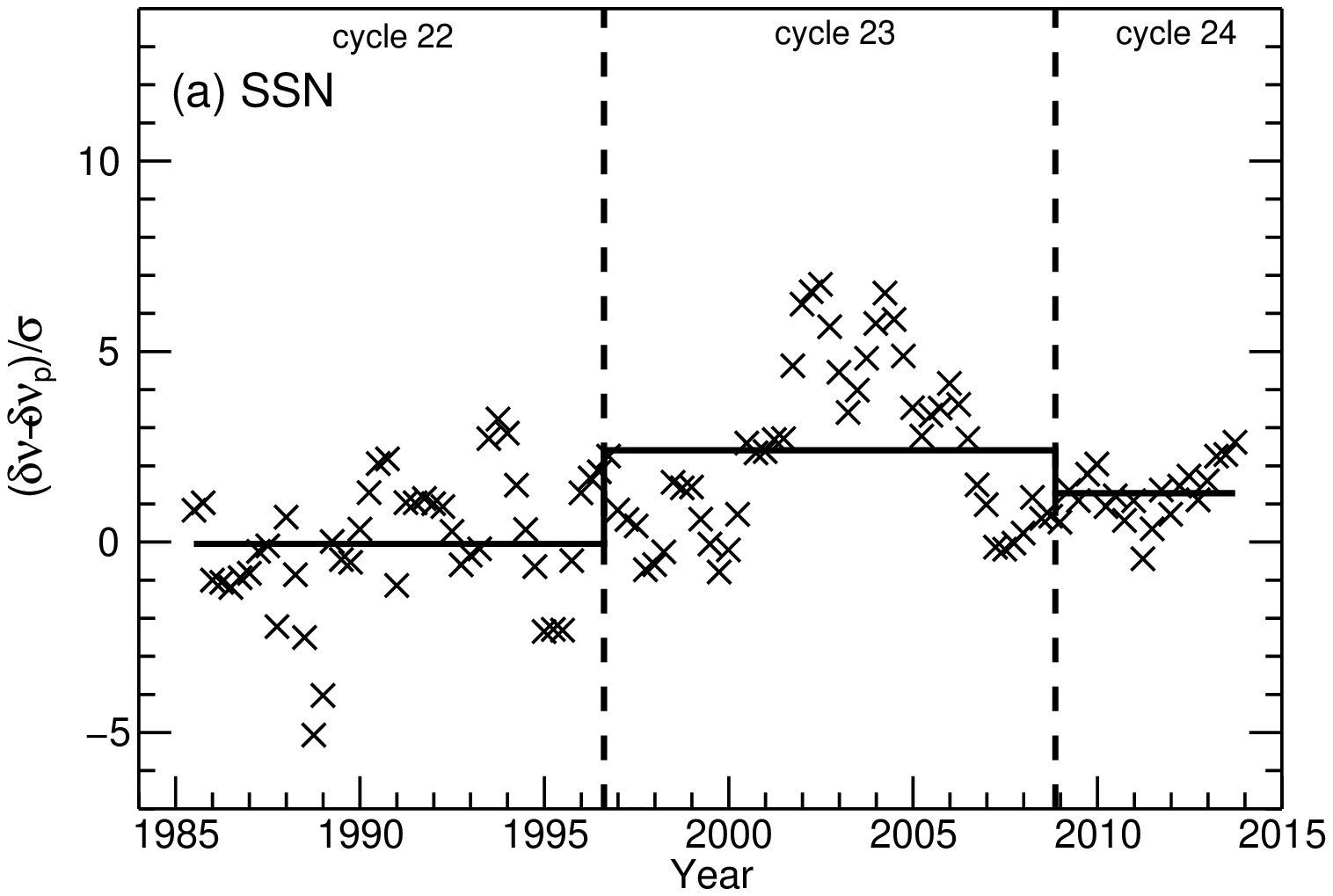}\hspace{1cm}
  \includegraphics[clip, width=0.45\textwidth]
{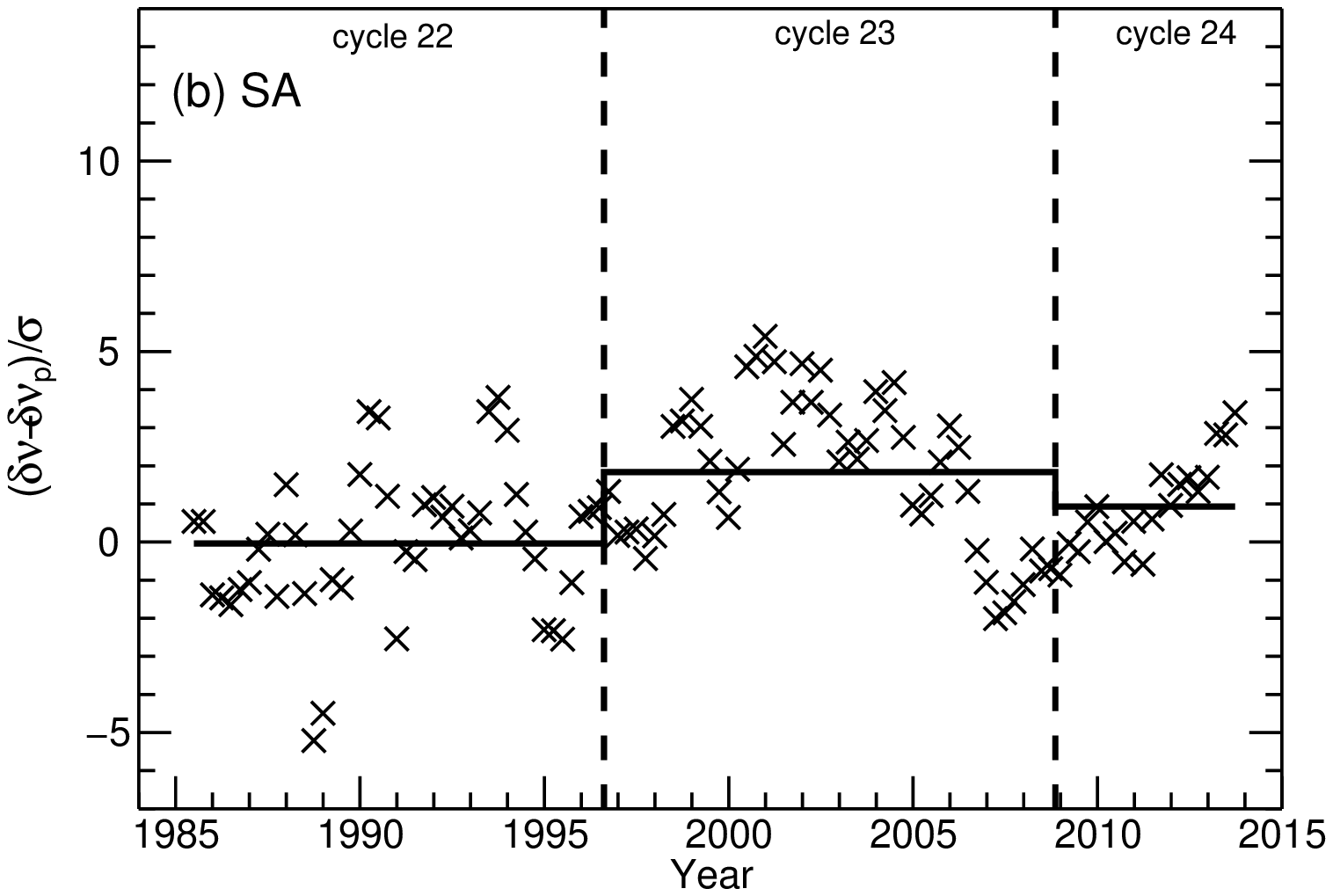}\\
  \includegraphics[clip, width=0.45\textwidth]
{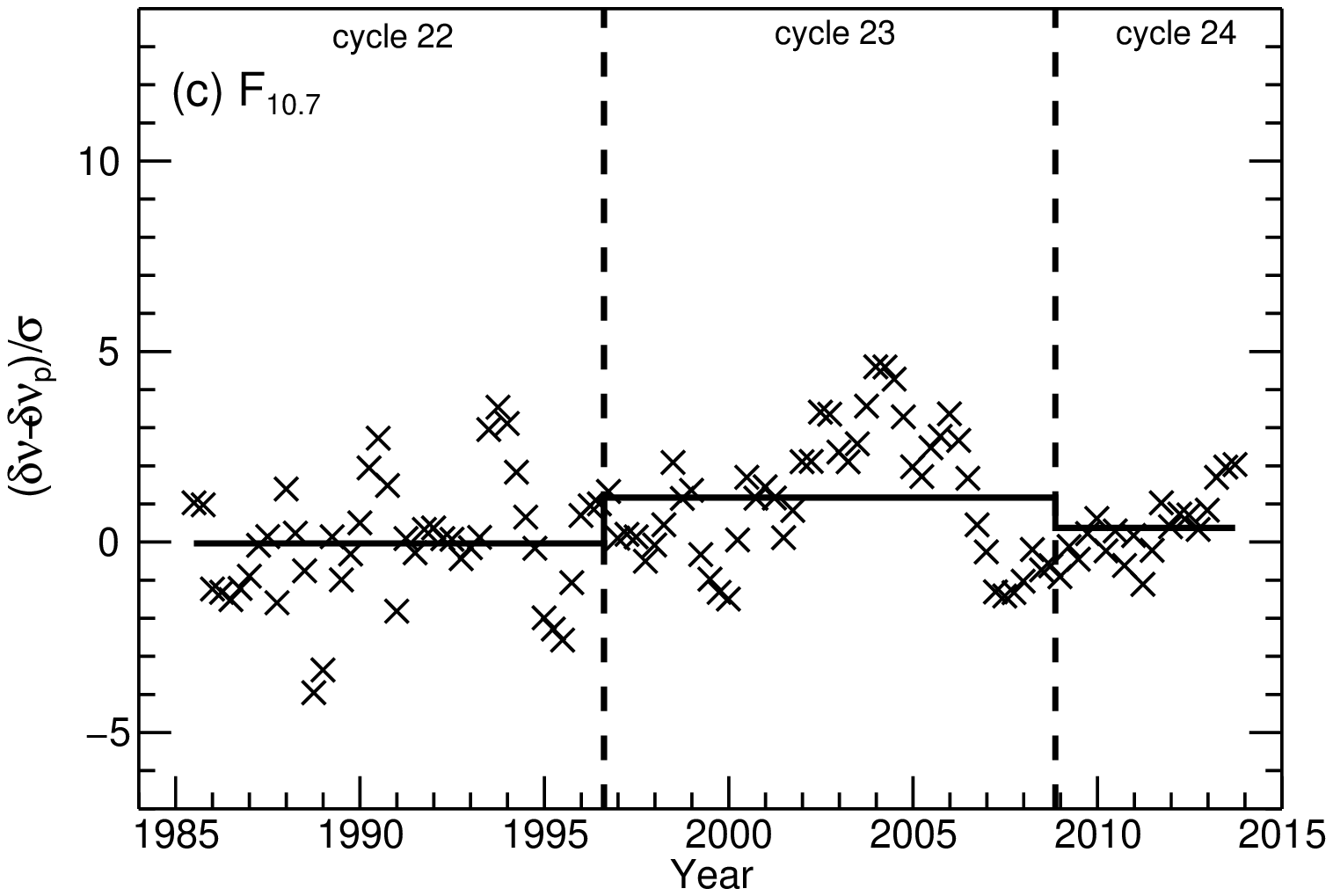}\hspace{1cm}
  \includegraphics[clip, width=0.45\textwidth]
{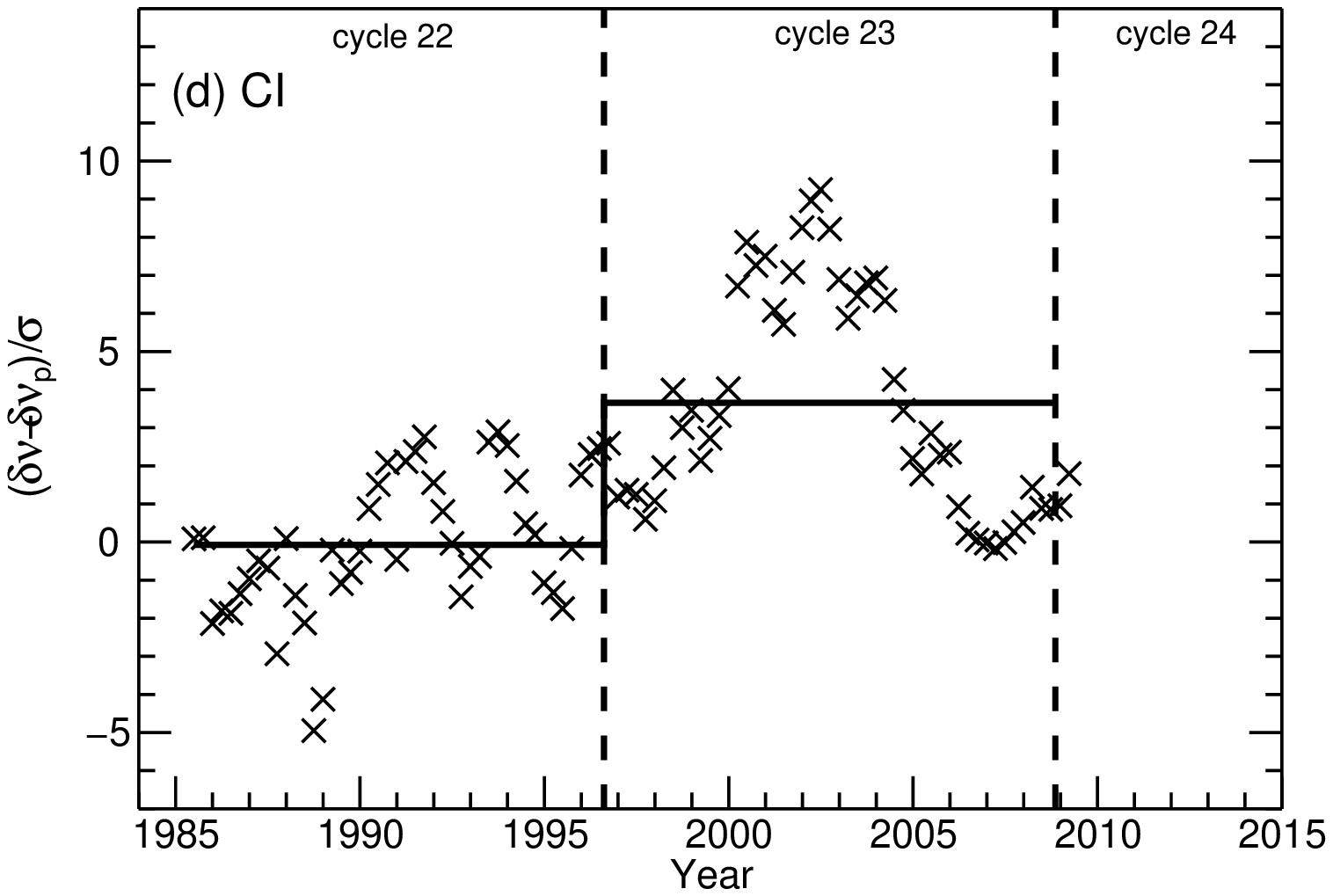}\\
  \includegraphics[clip, width=0.45\textwidth]
{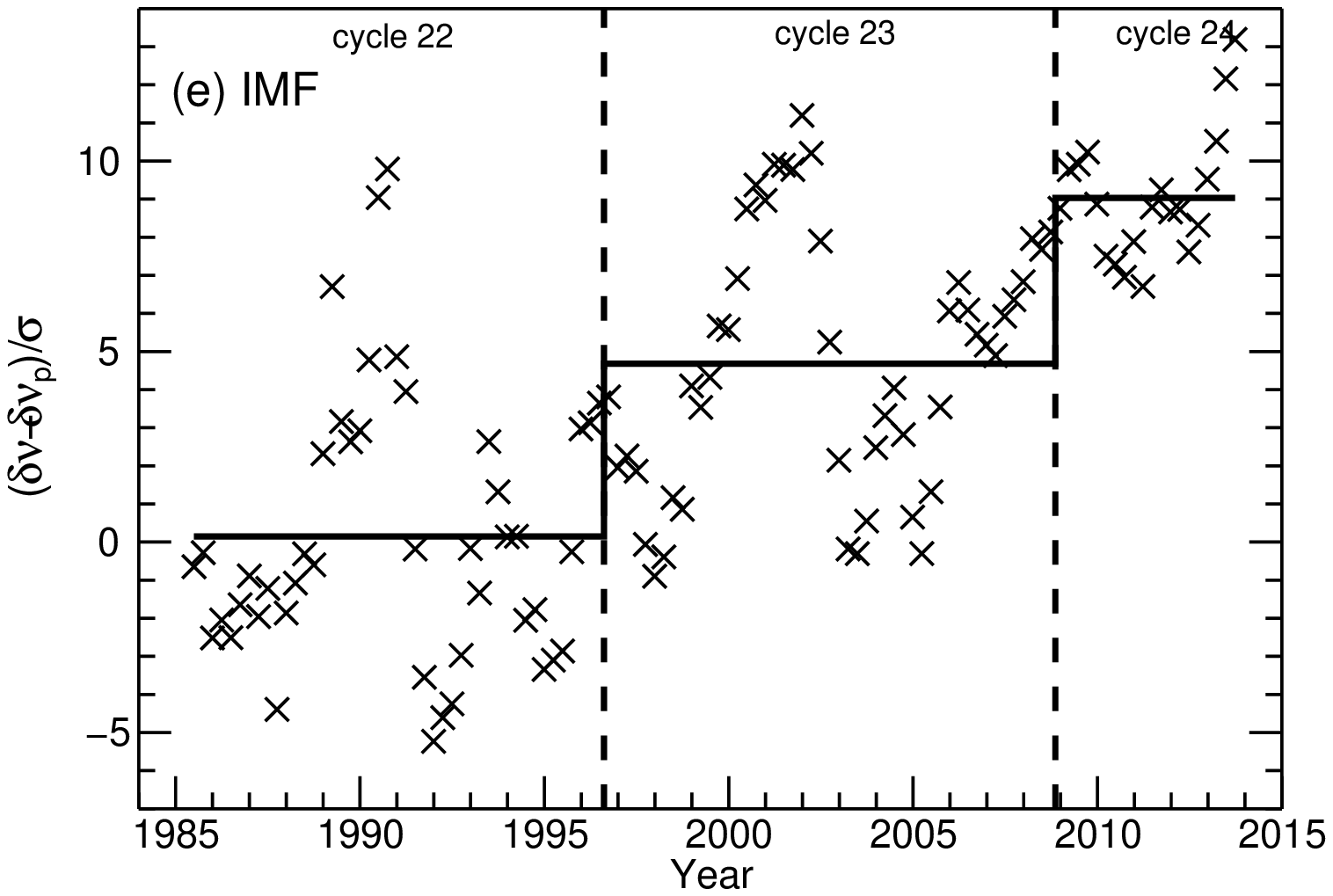}\hspace{1cm}
  \includegraphics[clip, width=0.45\textwidth]
{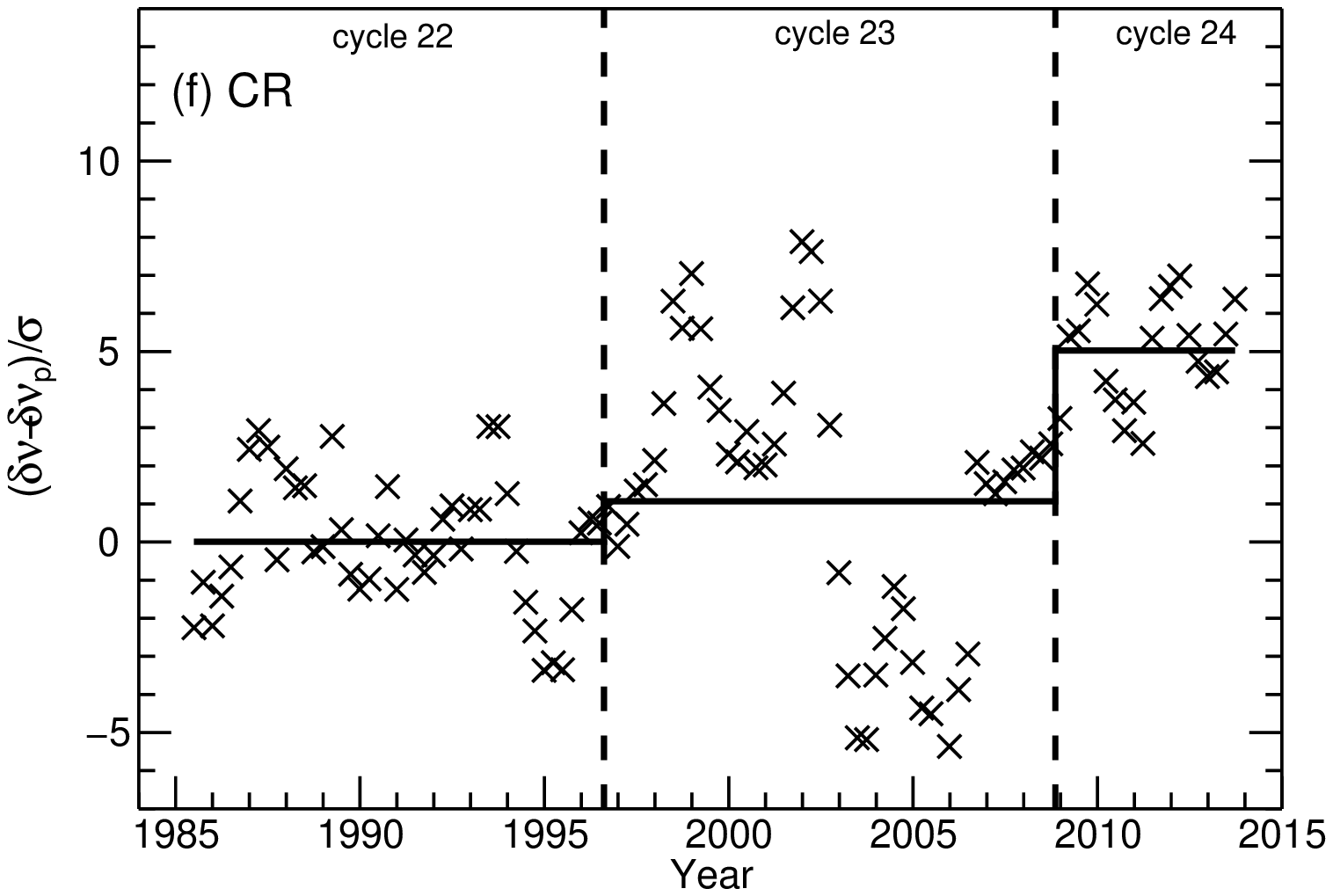}\\
  \caption{Scaled residuals $[(\delta\nu_i-\delta\nu_{\textrm{\scriptsize{p,}}i})/\sigma]$ as a 
function of time, where $\delta\nu_i$ is the observed helioseismic-frequency shift, $\sigma_i$ is 
the uncertainty associated with this frequency shift, and $\delta\nu_{\textrm{\scriptsize{p,}}i}$ is 
the frequency shift predicted by the linear regression for each proxy. The linear regressions were 
performed using data from Cycle 22 only and the values can be found in Table \ref{table[fits]}. The 
average values for each cycle are plotted as horizontal solid lines in black. Panel a: sunspot 
number. Panel b: sunspot area. Panel c: 10.7\,cm radio flux. Panel d: coronal index. Panel e: 
interplanetary magnetic field. Panel f: galactic cosmic-ray intensity. }
  \label{figure[residuals]}
\end{figure}

To assess the significance of these variations we determined the difference between the observed and 
predicted frequency shifts $[\delta\nu_i-\delta\nu_{\textrm{\scriptsize{p,}}i}]$ and then divided by 
the uncertainty associated with the frequency shifts $[\sigma_i]$. This uncertainty is derived from 
the uncertainty obtained from fitting the frequency power spectrum (described in Section \ref
{section[heliodata]}). These results are plotted in Figure \ref{figure[residuals]}. For convenience, 
the average ratio observed during each activity cycle is plotted. We remind the reader 
that the proxies were scaled with respect to the values observed in Cycle 22, it is therefore 
expected that the average ratio in Cycle 22 is approximately zero. For all proxies the agreement between the 
observed and  predicted frequency shifts is worse during Cycle 23 than Cycle 22. We note here that 
that does not necessarily mean that the agreement between the absolute values of the proxies and the 
helioseismic frequencies is worse as we must remember that we have used the relationship observed 
between the proxies and the frequencies in Cycle 22 to make the prediction. However, this figure 
does imply that the relationship between the proxies and the frequency shifts changes from Cycle 22 
to Cycle 23. On average the ratios are positive in Cycle 23, meaning that the predictions 
underestimate the shift in frequency. Furthermore, there are systematic trends in the ratios plotted 
in Figure \ref{figure[residuals]}, with the ratios being largest at solar maximum. This all implies 
that the linear regression is steeper in Cycle 23 than was observed in Cycle 22. This is supported 
by the results of Table \ref{table[fits]}, which shows that, for the solar proxies, the gradient of 
the linear regression is steeper in Cycle 23 than Cycle 22. We also note that the difference in the 
gradient is significant for SA and CI. Interestingly for the interplanetary proxies the 
gradient of the linear regression is steeper in Cycle 22 than Cycle 23. We remind the reader that CR is inversely scaled and so the gradient has a larger negative value in Cycle 22 than Cycle 23. Although on first inspection 
this seems contradictory to the result plotted in Figure \ref{figure[shifts_time]} it can be explained 
in terms of a change in the intercept (which acts as an offset). In particular this would produce 
better agreement between the proxies and the frequency shifts in the minimum between Cycles 23 and 
24 when large discrepancies are observed for the interplanetary proxies. We also note here that the 
reduced $\chi^2$ of the linear regressions of the interplanetary proxies is larger than for the 
solar proxies (with the exception of CR in Cycle 22, see Table \ref{table[fits]}).

The ratios plotted in Figure \ref{figure[residuals]} decrease in Cycle 24 compared to Cycle 23 for 
SSN, SA, and F$_{10.7}$, implying that the relationship between these proxies and the frequency shifts is 
returning to something similar to that observed in Cycle 22. On the other hand the ratios increase 
further for the interplanetary proxies (IMF and CR).

\begin{figure}
  \centering
  \includegraphics[clip, width=0.45\textwidth]{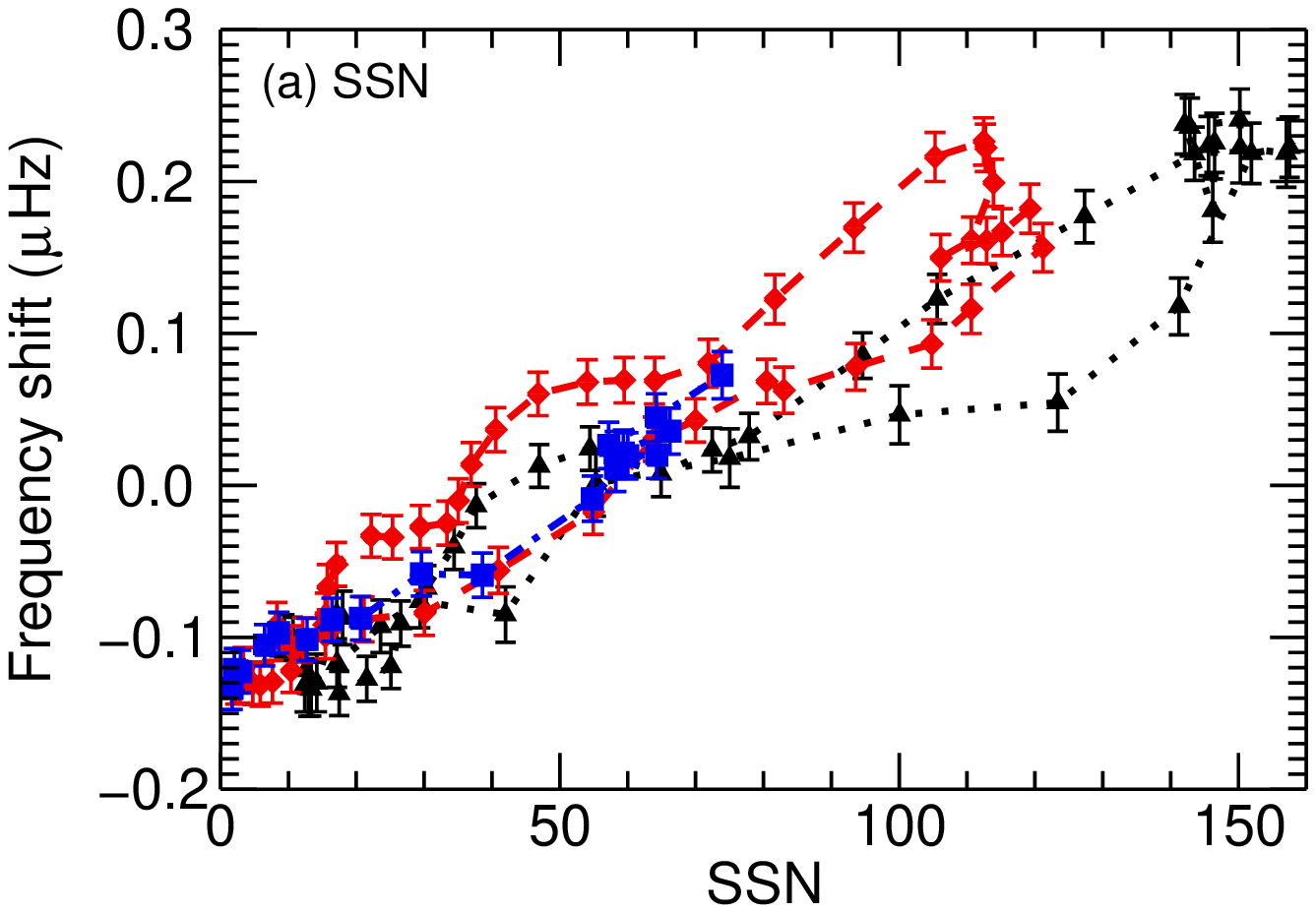}\hspace
{1cm}
  \includegraphics[clip, width=0.45\textwidth]{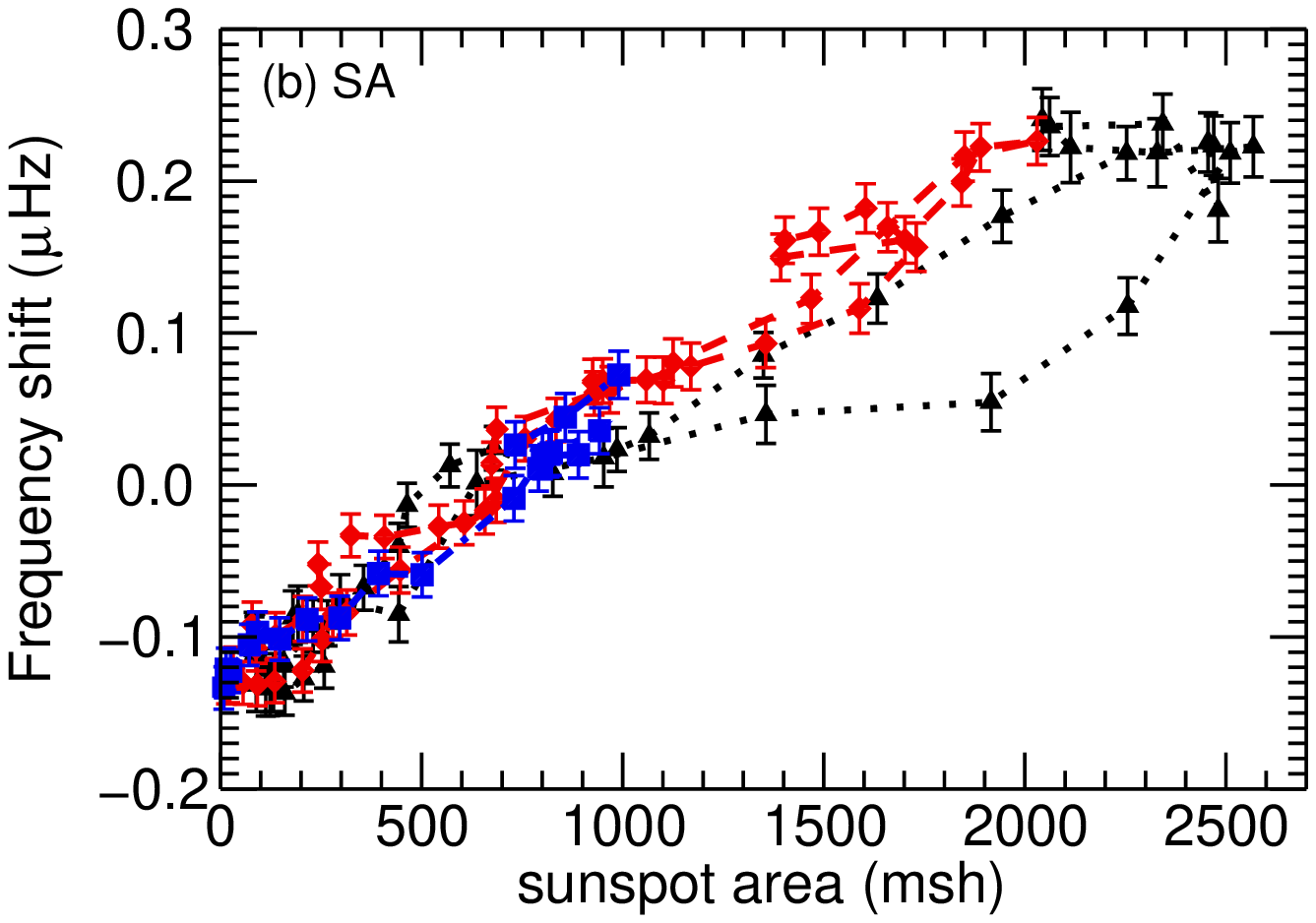}\\
  \includegraphics[clip, width=0.45\textwidth]{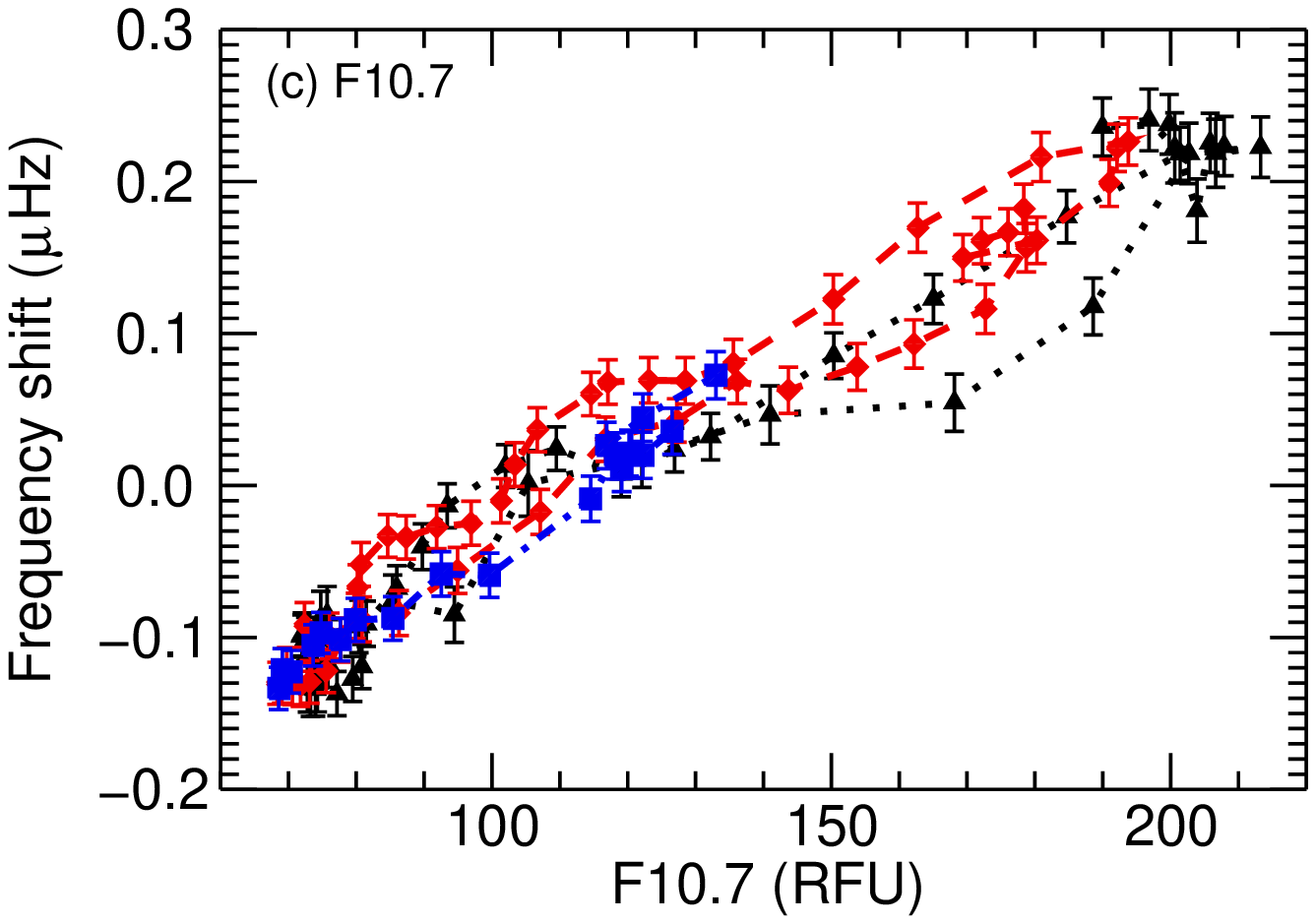}\hspace
{1cm}
  \includegraphics[clip, width=0.45\textwidth]{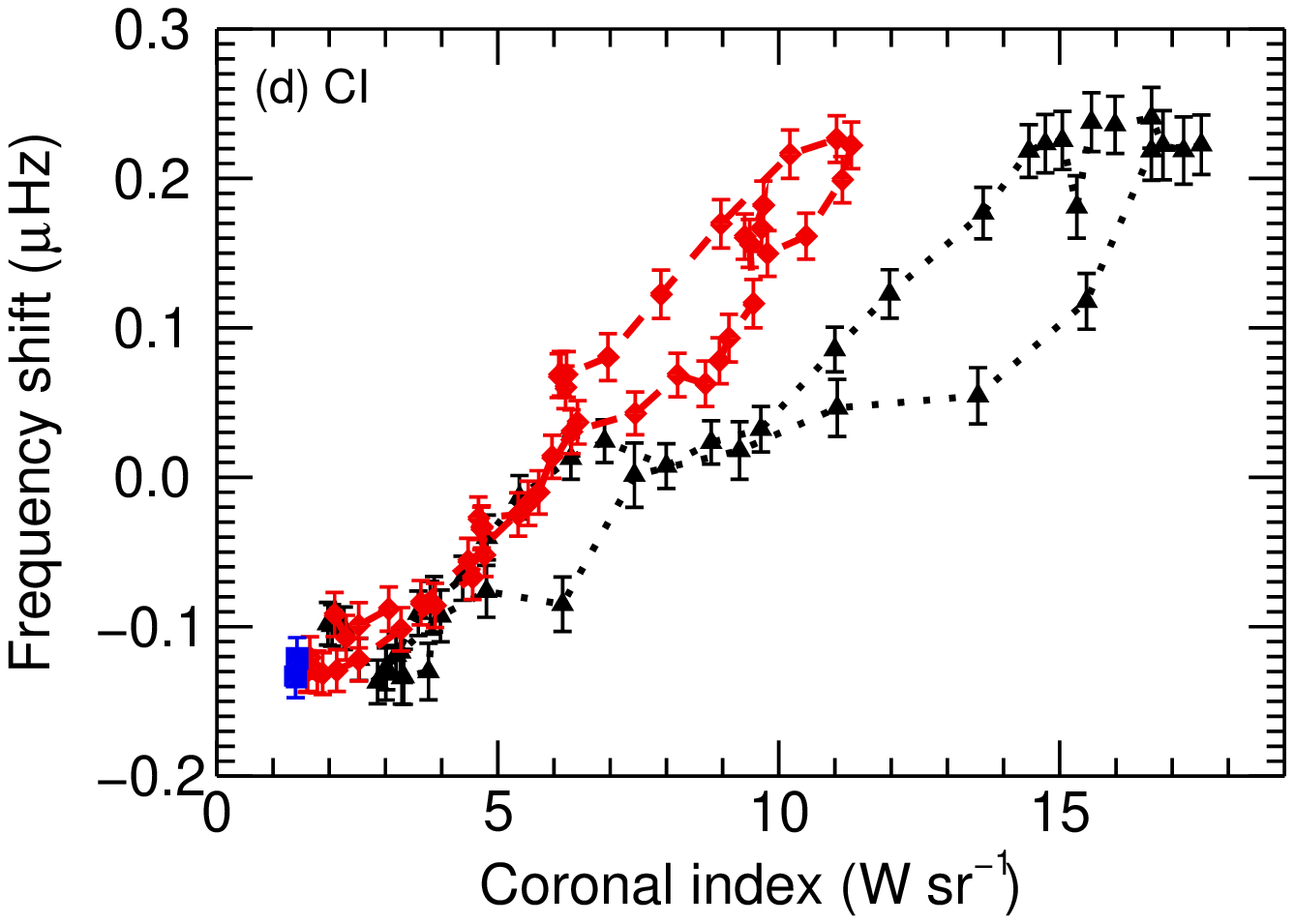}\\
  \includegraphics[clip, width=0.45\textwidth]{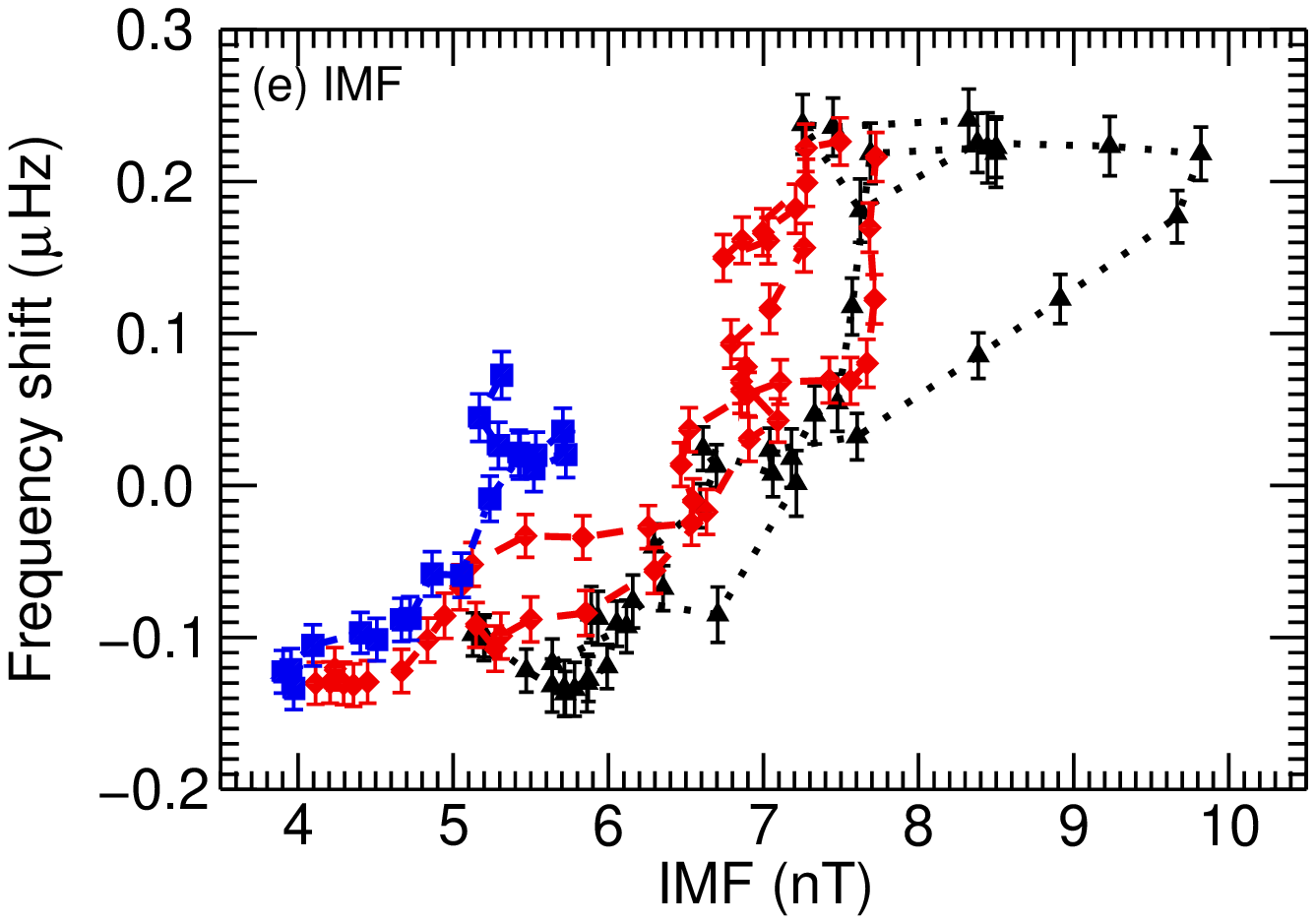}\hspace
{1cm}
  \includegraphics[clip, width=0.45\textwidth]{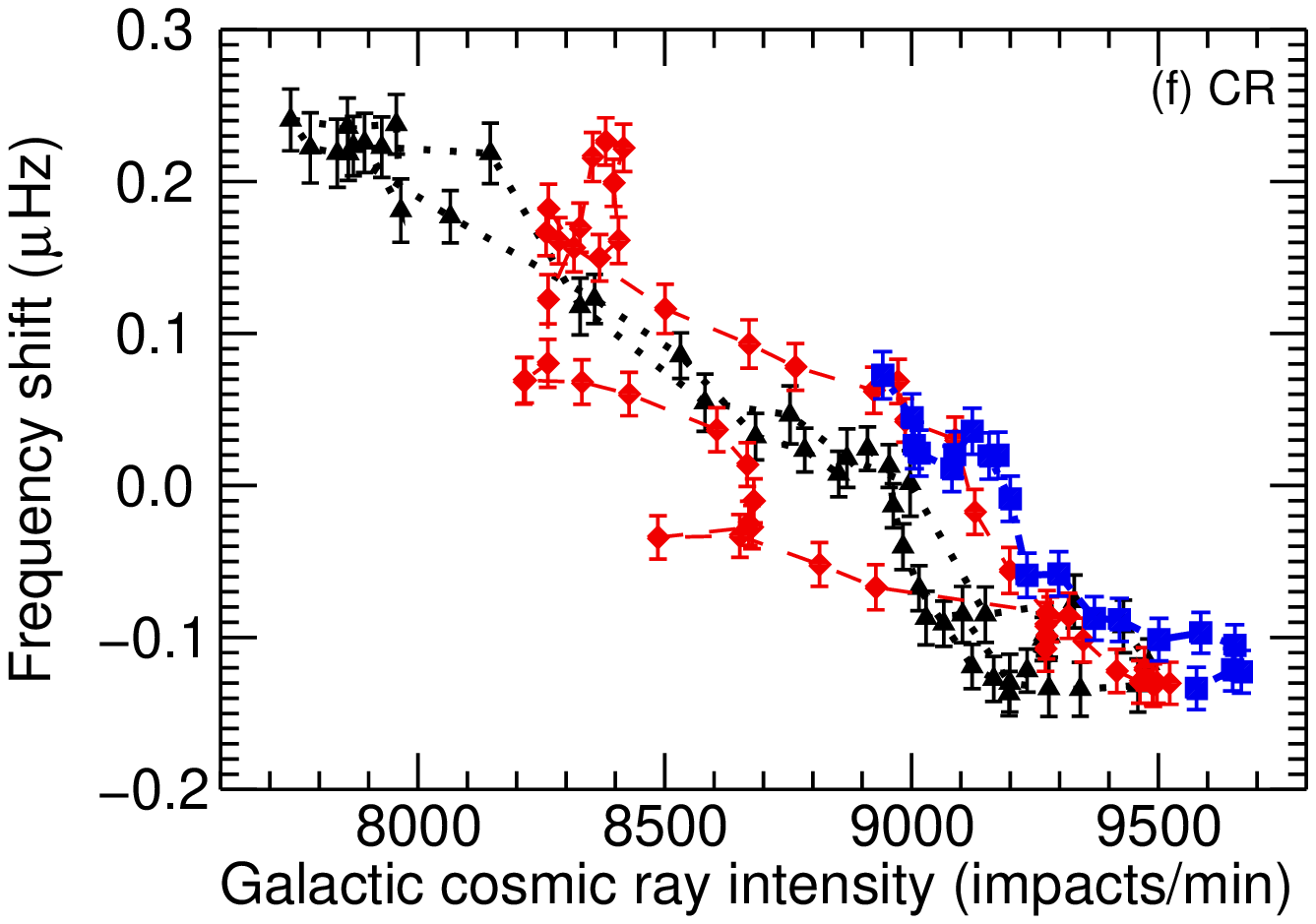}\\
  \caption{Frequency shifts as a function of the global proxies. The different colours represent the 
different cycles: the black-dotted line with triangle symbols represents Cycle 22, the red-dashed line with 
diamond symbols represents Cycle 23 and the blue-triple-dot-dashed line with square symbols represents 
Cycle 24.  Panel a: sunspot number. Panel b: sunspot area. Panel c: 10.7\,cm radio flux. Panel 
d: coronal index. Panel e: interplanetary magnetic field. Panel f: galactic cosmic-ray 
intensity.}
  \label{figure[shifts_vs_proxies]}
\end{figure}

Figure \ref{figure[shifts_vs_proxies]} shows the frequency shifts plotted directly as a function of 
the unscaled global activity proxies.  Since Figure \ref{figure[shifts_vs_proxies]} plots the unscaled 
activity proxies there is no reliance on the linear regression. However, Figure \ref{figure[shifts_vs_proxies]} still clearly shows that the relationship between the activity proxies and the 
helioseismic-frequency shifts varies from one cycle to the next. Notice that the relationship between 
the frequency shifts and the proxies is only approximately linear. The well-known hysteresis is also 
observed (\textit{e.g.} \opencite{1998A&A...329.1119J}; \opencite{2000JApA...21..357T}; \opencite
{2007ApJ...659.1749C}; \opencite{2009ApJ...695.1567J}). This occurs because of the distribution in 
latitude of the surface magnetic flux: Sunspots are known to appear at mid-latitudes and migrate 
towards the Equator as the cycle progresses. Some $p$-modes are more sensitive to low-latitude regions 
than others (depending on the harmonic degree $[\ell]$ and azimuthal degree $[m]$ that describe the 
spherical harmonic  of the mode) and so as the cycle evolves and the Sun's magnetic field migrates 
towards the Equator the influence of the Sun's magnetic field on each mode varies \citep{2000MNRAS.313..411M}. We note here that comparisons between solar-cycle associated frequency shifts 
and the appropriate spherical-harmonic decomposition of the surface magnetic flux result in a good 
linear relationship (\textit{e.g.} \opencite{1999ApJ...524.1084H}; \opencite{2002ApJ...580.1172H}).

To assess the significance of the hysteresis, we performed separate linear regressions for the rising 
and falling phases of each cycle and the coefficients can be found in Table \ref{table[fits_rise_fall]}. A comparison of the gradients of the rising and falling phases in any one cycle 
is very dependent on which proxy is considered and which cycle. For example, for any one cycle, 
the gradient of the linear regression obtained with the sunspot-area data is similar in the rising 
and falling phases, indicating a minimal amount of hysteresis. This is particularly true for Cycle 
23. On the other hand SSN, F$_{10.7}$, CI, and CR all show an increase in hysteresis in Cycle 23 compared 
with Cycle 22. The IMF shows a decrease in hysteresis in Cycle 23. We also note that the IMF is the 
only proxy for which the gradient is steeper in the rising phase of the cycle than the falling 
phase. However, we note from Figure \ref{figure[shifts_vs_proxies]}, and from the large reduced $
\chi^2$-values obtained that a linear relationship is not really appropriate. 

\begin{sidewaystable}\caption{Linear fits between helioseismic-frequency shifts and global proxies 
for rising and falling phases of the solar cycle. The units in which the gradient of the linear regression is measured 
depends on the proxy under consideration. We have used proxy unit (PU) as a global proxy unit where 
PU is equivalent to msh for SA; RFU for F$_{10.7}$; $\rm W\,sr^{-1}$ for CI; nT for IMF; impacts min$^{-1}$ for 
CR. We recall that SSN is a number with no units.\vspace{0.5cm}}\label{table[fits_rise_fall]}
\begin{tabular}{cc|ccc|ccc|cc}
  Proxy & Cycle & \multicolumn{3}{c}{Rising} &  \multicolumn{3}{c}{Falling} & \multicolumn{2}{c}
{Difference}\\
  & & Gradient  & Intercept  & Reduced & Gradient & Intercept & Reduced & Gradient & Intercept \\
  & & $(\times10^{-3})$  &  &  $\chi^2$ &  $(\times10^{-3})$ &  & $\chi^2$ &  & \\
  & & $[\rm\mu Hz\,PU^{-1}]$ & $[\rm\mu Hz]$ & & $[\rm\mu Hz\,PU^{-1}]$ & $[\rm\mu Hz]$ & & $\sigma$ & $\sigma$\\
  \hline
  \multirow{3}{*}{SSN} & 22 & $2.28\pm0.14$ & $-0.154\pm0.012$ & 2.71 & $2.51\pm0.12$ & $-
0.143\pm0.009$ & 2.63 & $1.2$ & $0.7$\\
   & 23 & $2.52\pm0.16$ & $-0.140\pm0.013$ & 1.42 &  $3.09\pm0.14$ & $-0.126\pm0.008$ & 2.31 & 
$2.7$ & $0.9$\\
   & 24 & $2.53\pm0.23$ & $-0.131\pm0.010$ & 0.575 & & & & & \\
  \hline
  \multirow{3}{*}{SA} & 22 & $0.139\pm0.008$ & $-0.132\pm0.011$ & 4.10 & $0.148\pm0.007$ & $-0.119\pm0.009$ & 3.32 & $0.8$ & $0.9$\\
   & 23 & $0.178\pm0.012$ & $-0.120\pm0.012$ & 2.14 & $0.182\pm0.008$ & $-0.127\pm0.008$ & 1.76 & 
$0.3$ & $0.5$\\
   & 24 & $0.181\pm0.017$ & $-0.128\pm0.010$ & 0.707 & & & & &  \\
  \hline
 \multirow{3}{*}{F10.7} & 22 & $2.46\pm0.15$ & $-0.298\pm0.019$ & 2.89 &  $2.55\pm0.13$ & $-
0.292\pm0.015$ & 1.20 & $0.5$ & $0.2$\\
   & 23 & $2.52\pm0.16$ & $-0.287\pm0.022$ & 1.20 & $2.81\pm0.12$ & $-0.302\pm0.014$ & 3.14 & 
$1.5$ & $0.6$\\
  & 24 & $2.87\pm0.26$ & $-0.324\pm0.265$ & 0.505 & & & & & \\
  \hline
  \multirow{2}{*}{CI} & 22 & $23.6\pm1.4$ & $-0.200\pm0.014$ & 2.27 &  $25.5\pm1.2$ & $-
0.176\pm0.011$ & 2.39 & $1.0$ & $1.3$\\
   & 23 & $33.9\pm2.2$ & $-0.192\pm0.016$ & 2.82 &  $38.7\pm1.7$ & $-0.207\pm0.011$ & 2.65 & 
$1.7$ & $0.8$\\
  \hline
 \multirow{3}{*}{IMF} & 22 & $128.6\pm7.7$ & $-0.870\pm0.052$ & 4.23 & $80.1\pm4.4$ & $-
0.540\pm0.030$ & 16.33 & $5.5$ & $5.5$ \\
   & 23 & $125.7\pm8.8$ & $-0.780\pm0.057$ & 11.27 & $82.9\pm4.0$ & $-0.493\pm0.024$ &11.27 & 
$4.4$ & $4.6$\\
  & 24 & $100.2\pm9.9$ & $-0.530\pm0.049$ & 3.62 & & & \\
  \hline
  \multirow{3}{*}{CR} & 22 & $-0.228\pm0.013$ & $2.02\pm0.12$ & 2.68 & $-0.249\pm0.012$ & 
$2.20\pm0.11$ & 3.11 & $1.2$ & $1.1$ \\
   & 23 & $-0.247\pm0.017$ & $2.28\pm0.15$ & 4.35 & $-0.209\pm0.011$ & $1.84\pm0.09$ & 17.98 & 
$1.9$ & $2.5$\\
  & 24 & $-0.258\pm0.024$ & $2.36\pm0.23$ & 2.10 & & & \\
  \hline
\end{tabular}
\end{sidewaystable}

\subsection{Short-term Variation}\label{section[2yr]}

\begin{figure}
  \centering
  \includegraphics[clip, width=0.45\textwidth]
{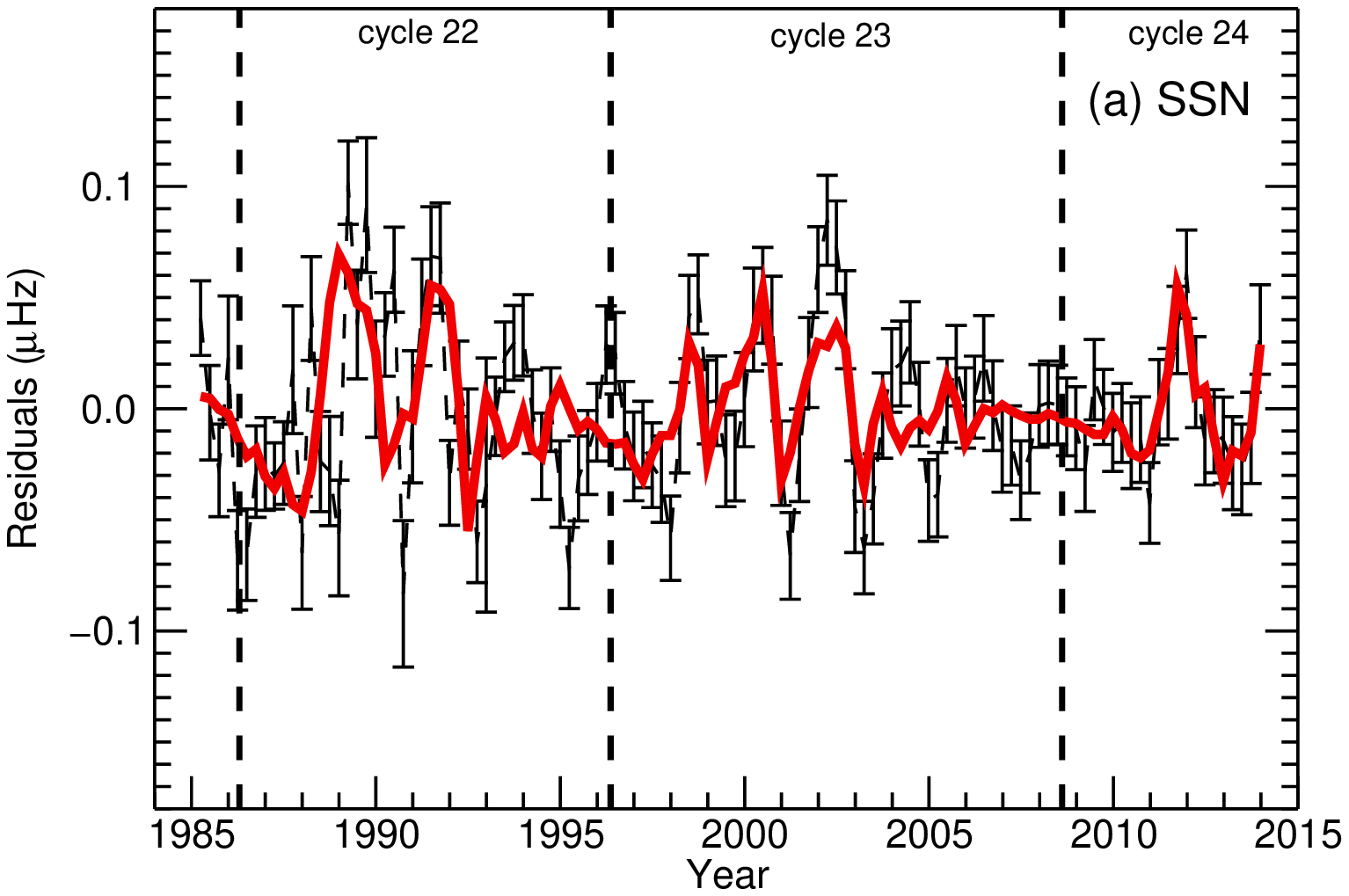}\hspace{1cm}
  \includegraphics[clip, width=0.45\textwidth]{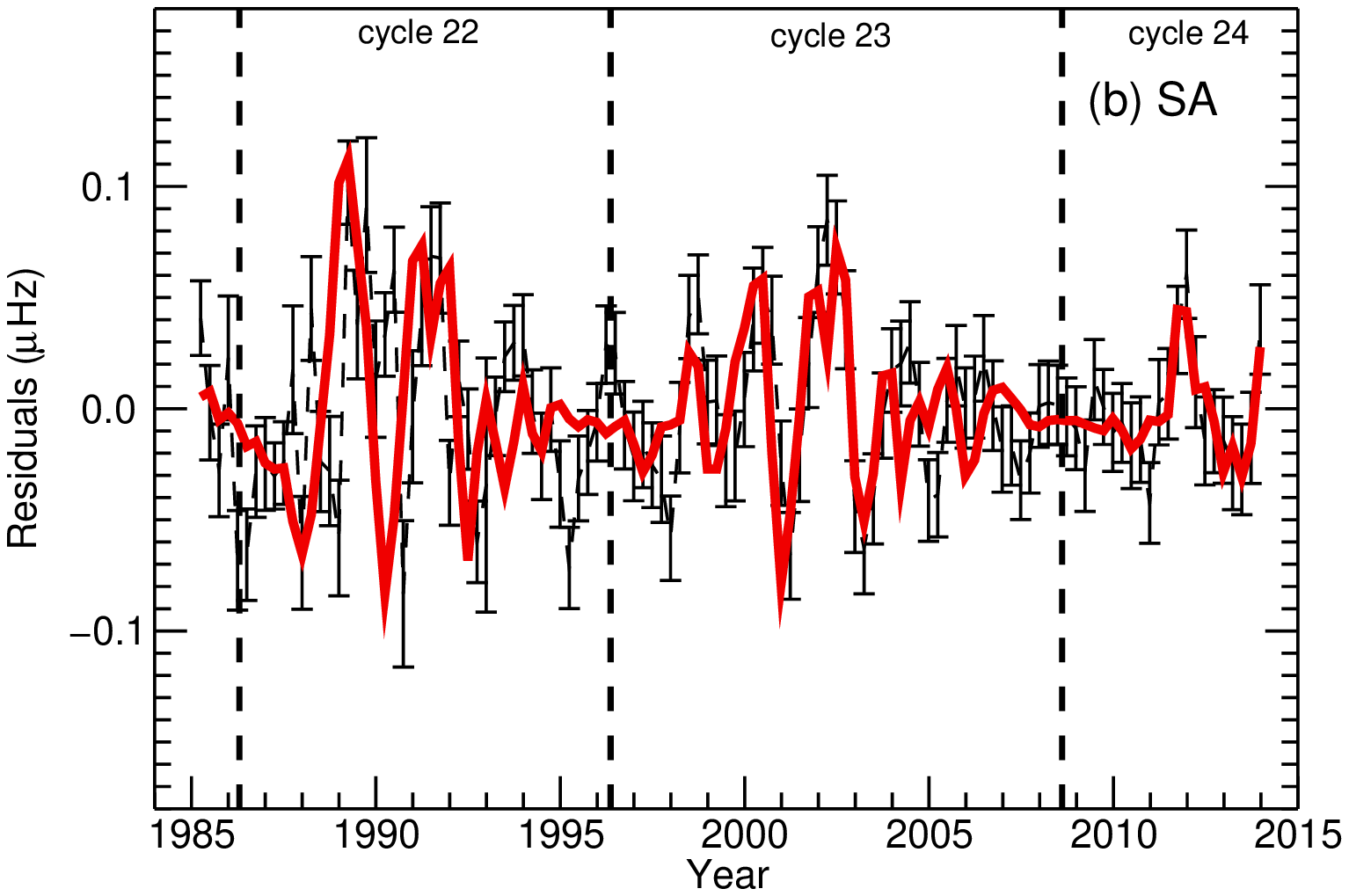}\\
  \includegraphics[clip, width=0.45\textwidth]
{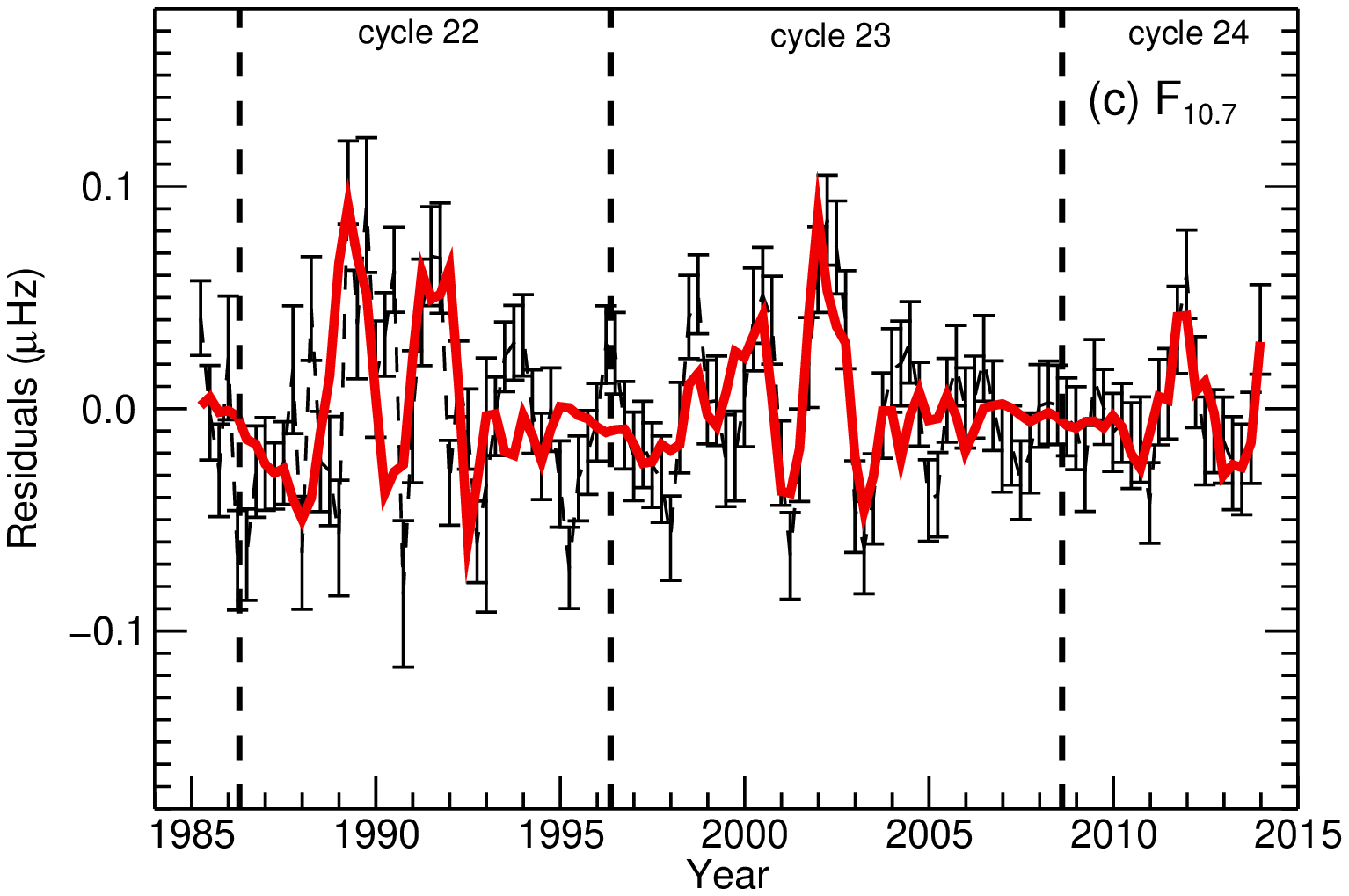}\hspace{1cm}
  \includegraphics[clip, width=0.45\textwidth]{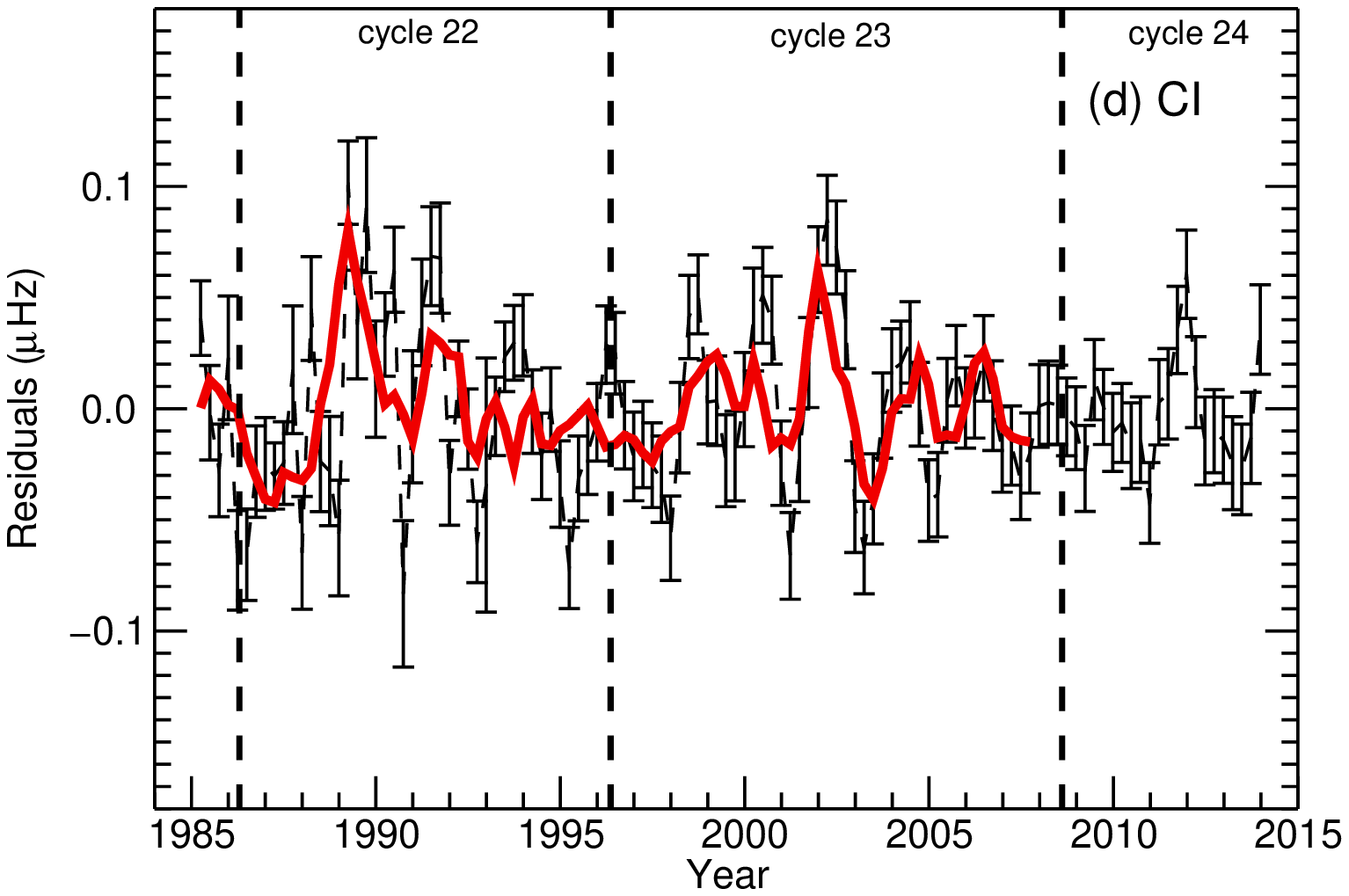}\\
  \includegraphics[clip, width=0.45\textwidth]
{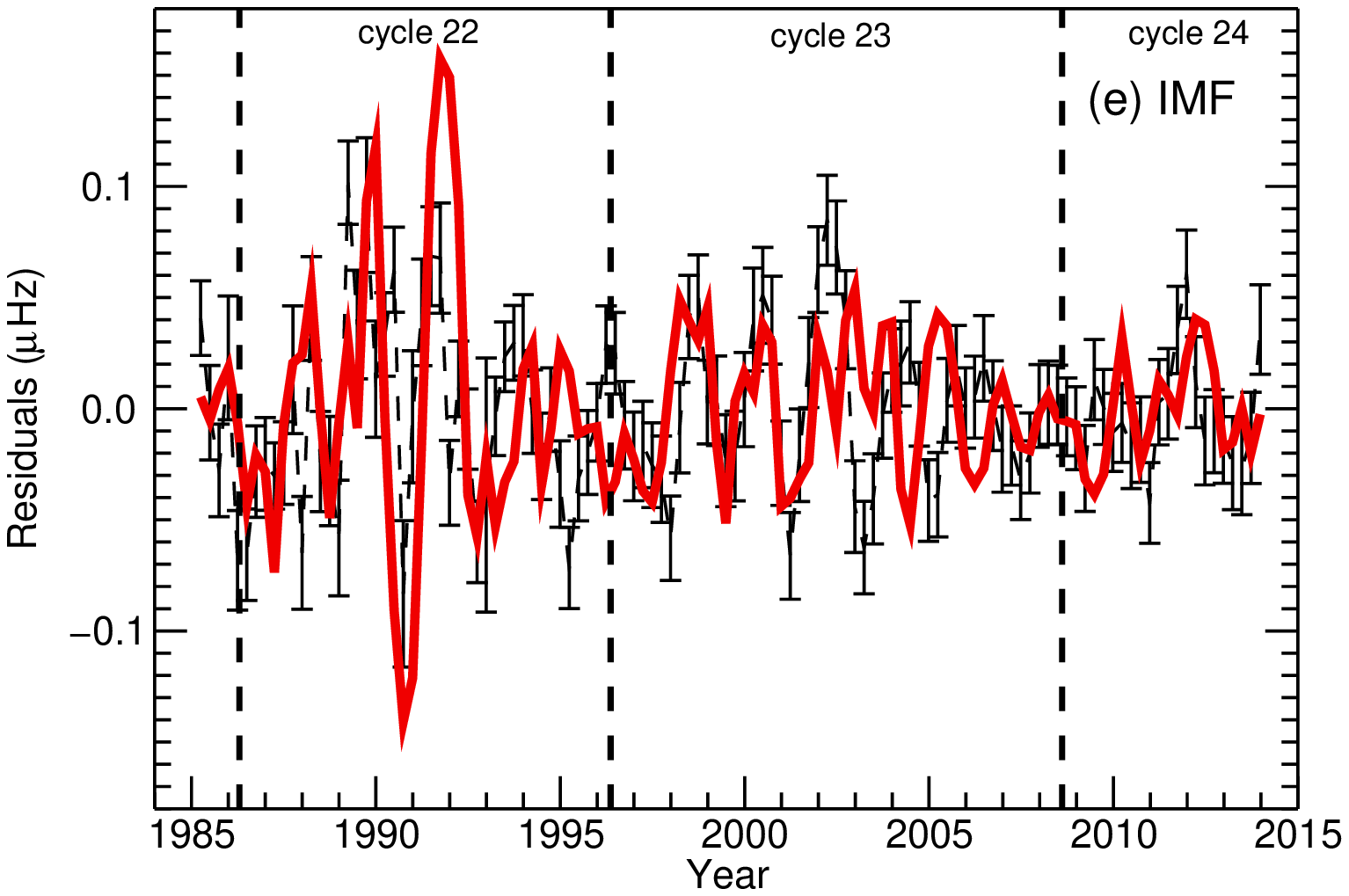}\hspace{1cm}
  \includegraphics[clip, width=0.45\textwidth]{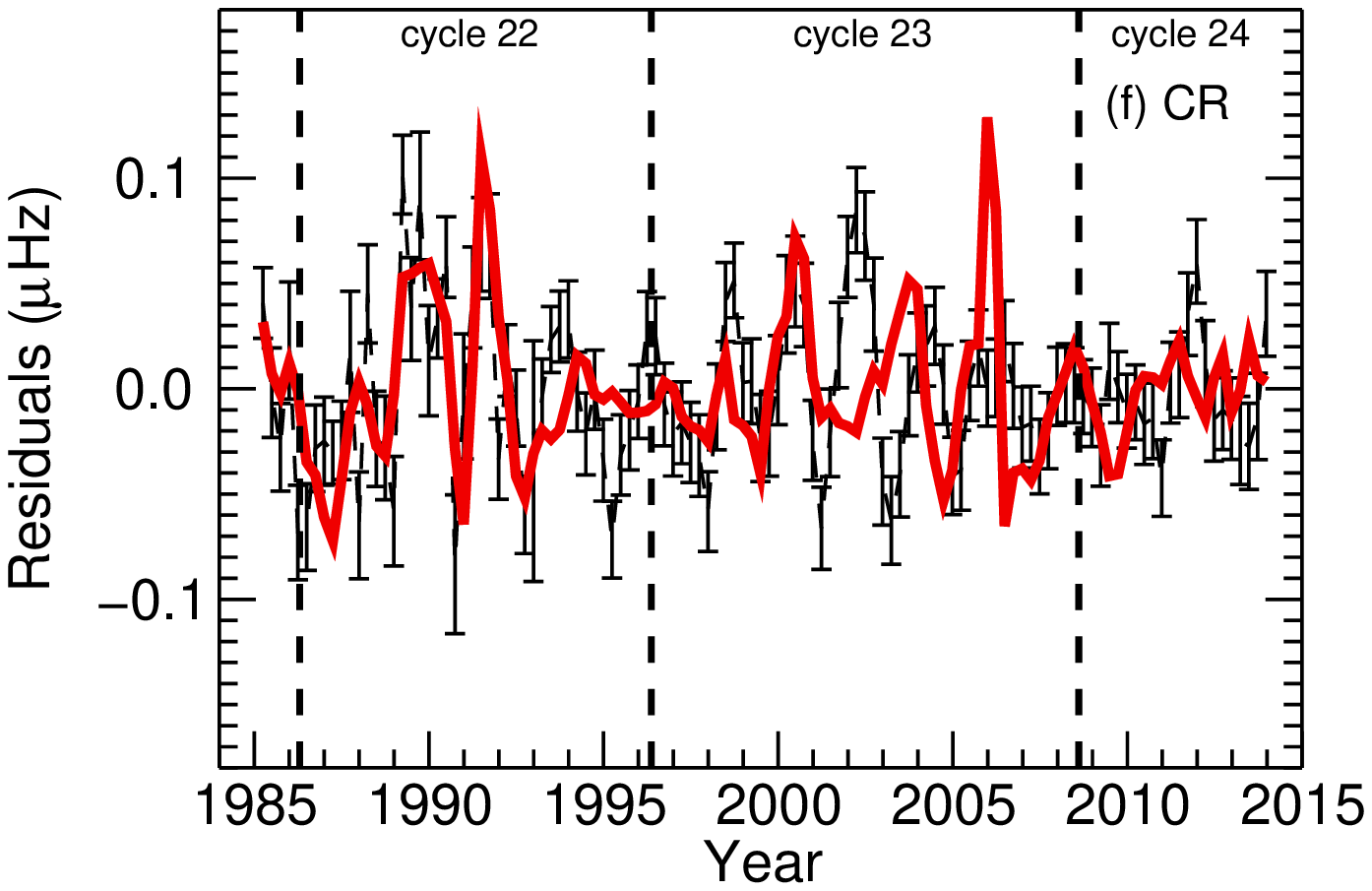}\\
  \caption{Frequency-shift residuals as a function of time (black-dashed line). The 
residuals are determined by subtracting three-year-smoothed frequency shifts from the frequency shifts 
plotted in Figure \ref{figure[shifts_time]}. Scaled versions of the activity proxies are plotted in 
coloured, solid lines. The activity proxies were scaled with respect to the raw proxies and 
frequency shifts and not the residuals themselves. As with Figure \ref{figure[shifts_time]} CR was 
regressed with a negative gradient. Panel a: scaled sunspot number. Panel b: scaled sunspot area. Panel c: 
scaled 10.7\,cm radio flux. Panel d: scaled coronal index. Panel e: scaled interplanetary magnetic field. Panel f: 
inversely scaled galactic cosmic ray intensity.}
  \label{figure[QBO]}
\end{figure}

It is well known that short term ($\approx$two-year) variations exist on top of the 11-year cycle in 
both the helioseismic-frequency shifts and other proxies of the solar activity (see \opencite
{2014SSRv..186..359B} for a recent review). Such variations are often referred to as a quasi-biennial oscillation (QBO). These variations can be seen in Figure \ref{figure[shifts_time]} but can 
be viewed more clearly once the 11-year cycle is removed. A smoothed version of the frequency shifts 
was subtracted from the raw frequency shifts (plotted in Figure \ref{figure[shifts_time]}), where the 
smoothing was performed over three years. At the edges additional end points based on the values of the first and last values respectively were used to artificially extend the array and allow the smoothing to be completed. The residuals are plotted in Figure \ref{figure[QBO]}. Once again 
a scaled version of the proxies is plotted for comparison (note again that CR is inversely scaled). To maintain consistency, the linear 
regressions were performed using the raw proxies and frequency shifts, not the residuals. In Figure 
\ref{figure[QBO]} the short-term variations are clear in both the frequency shifts and the proxies, 
especially around solar maximum. An enhanced amplitude around solar maximum is a well-known property 
of the QBO  (see \opencite{2014SSRv..186..359B} and references therein). Note that the amplitude of 
the QBO is consistent between the proxies and the frequency shifts despite the fact that the raw 
values, which are dominated by the 11-year cycle, were used to scale the proxies. 

Table \ref{table[QBO]} shows the correlations of the unscaled proxy residuals with the frequency -shift residuals. We note that the correlations were determined using only independent data points (here those with even subscripts were used). There is a definite split between the solar proxies (SSN, SA, F$_{10.7}$, and CI) and 
the interplanetary proxies (IMF, CR). For the solar proxies, the correlation is higher in Cycle 23 
than in Cycle 22. However, the poor correlations observed during Cycle 22 could simply be a 
reflection of the fact that the helioseismic data were of poorer quality. The situation is different 
for the interplanetary proxies, where the correlations are lower in Cycle 23 (and Cycle 24) than was 
observed in Cycle 22. We must remember that we are currently in the maximum phase of Cycle 24, when the amplitude of the quasi-biennial periodicity is at a maximum and so it would be misleading to compare the correlation observed in Cycle 24 with 
Cycles 22 and 23. However, we note from Figure \ref{figure[QBO]} and Table \ref{table[QBO]} that the 
data are highly correlated with the solar proxies in Cycle 24 thus far. As one might expect, correlations between activity proxies and helioseismic-frequency shifts (not residuals) are higher in the rising and falling phases, where relatively large changes in time are observed, than around solar maxmimum and minimum, where the proxies and helioseismic frequencies are approximately constant (\textit{e.g.} \opencite{2009ApJ...695.1567J}). The QBO, on the other hand, appears to be most highly correlated around solar maximum, where the amplitude of the QBO is largest. 

\begin{table*}\caption{Correlations between helioseismic-frequency residuals and solar activity proxies. We note that only the rising phase and a part of the maximum of Cycle 24 are analysed.}\label{table[QBO]}
\begin{tabular}{cc|cc|cc}
  \hline
 % % after \\: \hline or \cline{col1-col2} \cline{col3-col4} ...
  Proxy & Cycle & \multicolumn{2}{c}{Pearson's} &\multicolumn{2}{c}{Spearman's} \\
  \cline{3-4}\cline{5-6}
   & & $r$ & Probability & $r_s$ & Probability \\
  \hline
  \multirow{4}{*}{SSN} & all & 0.57 & $6.1\times10^{-7}$ & 0.47 & $1.6\times10^{-4}$ \\
   & 22 & 0.45 & 0.014 & 0.31 & 0.15 \\
   & 23 & 0.77 & $1.8\times10^{-6}$ & 0.63 & $9.0\times10^{-4}$ \\
   & 24 & 0.90 &$7.1\times10^{-6}$ & 0.78 & 0.0030 \\
  \hline
  \multirow{4}{*}{SA} & all & 0.47 & $8.1\times10^{-5}$ & 0.32 & 0.014 \\
   & 22 & 0.37 & 0.040 & 0.22 & 0.32 \\
   & 23 & 0.57 & 0.0015 & 0.45 & 0.028 \\
   & 24 & 0.90 & $4.1\times10^{-6}$ & 0.77 & 0.0034 \\
  \hline
  \multirow{4}{*}{F$_{10.7}$} & all & 0.64 & $8.3\times10^{-9}$ & 0.50 & $6.3\times10^{-5}$ \\
  & 22 & 0.55 & 0.0028 & 0.32 & 0.14 \\
  & 23 & 0.76 & $2.6\times10^{-6}$ & 0.63 & $9.6\times10^{-4}$ \\
  & 24 & 0.90 & $4.18\times10^{-6}$ & 0.76 & 0.0045 \\
  \hline
  \multirow{3}{*}{CI} & all & 0.42 & $4.3\times10^{-4}$ & 0.34 & 0.065 \\
   & 22 & 0.50 & 0.0074 & 0.39 & 0.065 \\
   & 23 & 0.67 & $1.3\times10^{-4}$ & 0.65 & $8.6\times10^{-4}$ \\
  \hline
  \multirow{4}{*}{IMF} & all & 0.52 & $7.6\times10^{-6}$ & 0.42 & $8.6\times10^{-4}$ \\
   & 22 & 0.62 & $5.7\times10^{-4}$ & 0.59 & 0.0029 \\
   & 23 & 0.30 & 0.082 & 0.22 & 0.31 \\
   & 24 & 0.33 & 0.15 & 0.54 & 0.071 \\
  \hline
  \multirow{3}{*}{CR} & all & -0.45 & $1.4\times10^{-4}$ & -0.40 & 0.0016 \\
   & 22 & -0.70 & $4.5\times10^{-5}$ & -0.67 & $5.2\times10^{-4}$ \\
   & 23 & -0.12 & 0.30 & -0.12 & 0.56 \\
   & 24 & -0.11 & 0.37 & -0.12 & 0.7 \\
  \hline
\end{tabular}
\end{table*}

\section{Discussion}\label{section[discussion]}
If proven to be true, one of the consequences of the apparent decrease in the average magnetic field 
of sunspots observed by, for example, \inlinecite{2012ApJ...757L...8L} is that, if the decrease 
continues, fewer small sunspots will be observed. \inlinecite{2006ApJ...649L..45P} found that 
sunspots can only form if the magnetic-field strength exceeds 1500\,G. If the magnetic-field 
strength is weaker, the field penetrating the solar surface is more diffuse rather than clustering in 
small pores and spots. Not only does this explain the apparent divergence 
between the 10.7\,cm flux (which is sensitive to both strong magnetic field regions and weaker 
faculae and plages) and the SSN \citep{2012ApJ...757L...8L} but it is also consistent with the 
apparent deficit in the number of small spots observed at the maximum of Cycle 23 compared to Cycle 
22 \citep{2011ApJ...731...30K, 2011A&A...536L..11L, 2012JSWSC...2A..06C}. Sunspot area is dominated 
by the largest active regions and so is less sensitive to this small-spot deficit than SSN, which 
gives equal weighting to spots of all sizes and is, therefore, dominated by small spots, which are 
most abundant. The coronal index, which  is connected with the green-line emission produced by 
ionised iron, is associated with a specific temperature of the plasma. The discrepancies with the 
other activity proxies show that, while the coronal index may be affected by the sunspot deficit, 
other features, such as coronal holes, are very important in determining the observed values. For 
example \inlinecite{2010ApJ...712..813A} observed that the area of low-latitude coronal holes 
increased in the declining phase of Cycle 23. 

These results therefore imply that the observed frequency shifts are the result of both strong and 
weak-magnetic-field regions. Although the size of the frequency shift is dependent on the strength 
of the magnetic field, a shift of some magnitude will be observed no matter what the strength of the 
magnetic field is. This is similar to the 10.7\,cm flux but in contrast to the SSN, which has a 
magnetic-field strength threshold, below which no sunspot is formed. It is, therefore, 
understandable that better agreement is observed between the helioseismic frequencies and the 
10.7\,cm flux than the SSN. This is in agreement with the results of \inlinecite
{2007ApJ...659.1749C}. Given the frequency range under consideration here, our results are also in agreement with \inlinecite{2015arXiv150207607S}, who found that the variation in the frequencies of high-frequency modes reflects the variation observed in the 10.7\,cm radio flux. These results also imply that the additional effects that brought about a 
reduction in the CI in Cycle 23 (such as low-latitude coronal holes) are not reliant on the magnetic 
field in the solar interior. There is also clear evidence for a change in the relationship between 
the interplanetary proxies and the helioseismic data since Cycle 22.

The differences in the 11-year cycles become even more interesting when we consider the QBO. For the 
solar proxies, the frequency-shift residuals and the proxy residuals are highly correlated in all 
cycles, and in particular Cycles 23 and 24. Additionally, there is no divergence in amplitude 
indicating that, in terms of the QBO, there is no change in relationship between the proxies and the 
helioseismic-frequency shifts. Conversely the interplanetary-proxy residuals and the helioseismic-frequency-shift residuals are poorly correlated in Cycles 23 and 24. Interestingly, the decoupling of the 
QBOs appears to have occurred around 2000 (see Figure \ref{figure[QBO]}). This is approximately the 
same time that the 11-year cycles in the global proxies began to diverge. 

Finally we recall that in Figure \ref{figure[residuals]}, for the solar proxies, the scaled residuals 
decreased in Cycle 24 compared to Cycle 23. This raises the tantalising question as to whether we 
are simply observing a 22-year Hale cycle effect, where by the frequency shifts and the proxies 
display one relationship in even cycles and a different relationship in odd cycles. 

%%%%%%%%%%%%%%%%%%%%%%%%%%%%%%%%%%%%%%%%%%%%%%%%%%%%%%%%%%%%%%%%%%%%%%%%%%%
%% Acknowledgements
%
\begin{acks}
A.-M. Broomhall thanks the Institute of Advanced Study, University of Warwick for their support. V.M. Nakariakov: This work 
was supported by the European Research Council under the \textit{SeismoSun} Research Project No. 
321141, STFC consolidated grant ST/L000733/1, and the BK21 plus program through the National 
Research Foundation funded by the Ministry of Education of Korea. We thank the Birmingham 
Solar Oscillations Network, IZMIRAN Cosmic Ray Group, NOAA NGDC, OMNIWeb, and the Royal Observatory 
(Greenwich) for making their data freely available. We acknowledge use of NASA/GSFC's Space Physics 
Data Facility's OMNIWeb (or CDAWeb or ftp) service, and OMNI data. We acknowledge the Leverhulme 
Trust for funding the ``Probing the Sun: inside and out'' project upon which this research is based. The research leading to these results has received funding from the European Community's Seventh Framework Programme ([FP7/2007-2013]) under grant agreement n° 312844 (see Article II.30. of the Grant Agreement).
\end{acks}

\section*{Disclosure of Potential Conflicts of Interest} The authors declare that they have no conflicts of interest.

\bibliographystyle{spr-mp-sola}
\bibliography{comparison} 

\begin{thebibliography}{51}
% BibTex style file: spr-mp-sola.bst (nameyear), 2015-03-09
\ifx\bisbn     \undefined \def\bisbn  #1{ISBN #1}\fi
\ifx\binits    \undefined \def\binits#1{#1}\fi
\ifx\bauthor   \undefined \def\bauthor#1{#1}\fi
\ifx\batitle   \undefined \def\batitle#1{#1}\fi
\ifx\bjtitle   \undefined \def\bjtitle#1{\textit{#1}}\fi
\ifx\bvolume   \undefined \def\bvolume#1{\textbf{#1}}\fi
\ifx\byear     \undefined \def\byear#1{#1}\fi
\ifx\bissue    \undefined \def\bissue#1{#1}\fi
\ifx\bfpage    \undefined \def\bfpage#1{#1}\fi
\ifx\blpage    \undefined \def\blpage #1{#1}\fi
\ifx\burl      \undefined \def\burl#1{\textsf{#1}}\fi
\ifx\href      \undefined \def\href#1#2{\textsf{#2}}\fi
\ifx\betal     \undefined \def\betal{\textit{et al.}}\fi
\ifx\bctitle   \undefined \def\bctitle#1{#1}\fi
\ifx\beditor   \undefined \def\beditor#1{#1}\fi
\ifx\bbtitle   \undefined \def\bbtitle#1{\textit{#1}}\fi
\ifx\bedition  \undefined \def\bedition#1{#1}\fi
\ifx\bseriesno \undefined \def\bseriesno#1{\textbf{#1}}\fi
\ifx\blocation \undefined \def\blocation#1{#1}\fi
\ifx\bsertitle \undefined \def\bsertitle#1{\textit{#1}}\fi
\ifx\bsnm      \undefined \def\bsnm#1{#1}\fi
\ifx\bsuffix   \undefined \def\bsuffix#1{#1}\fi
\ifx\bparticle \undefined \def\bparticle#1{#1}\fi
\ifx\barticle  \undefined \def\barticle#1{}\fi
\ifx\binstitute  \undefined \def\binstitute#1{#1}\fi
\ifx\bpublisher  \undefined \def\bpublisher#1{#1}\fi
\ifx\doiurl    \undefined
  \def\doiurl#1{\href{http://dx.doi.org/#1}{\textsf{DOI}}}\fi
\ifx\arxivurl  \undefined
  \def\arxivurl#1{\href{http://arxiv.org/abs/#1}{\textsf{arXiv}}}\fi
\ifx\adsurl    \undefined
  \def\adsurl#1{\href{http://adsabs.harvard.edu/abs/#1}{\textsf{ADS}}}\fi
\ifx\botherref \undefined \def\botherref#1{}\fi
\ifx\url       \undefined \def\url#1{\textsf{#1}}\fi
\ifx\bchapter  \undefined \def\bchapter#1{}\fi
\ifx\bbook     \undefined \def\bbook#1{}\fi
\ifx\bcomment  \undefined \def\bcomment#1{#1}\fi
\ifx\oauthor   \undefined \def\oauthor#1{#1}\fi
\ifx\citeauthoryear \undefined\def \citeauthoryear#1{#1}\fi
\ifx\endbibitem\undefined \def\endbibitem{}\fi
\ifx\bconflocation  \undefined \def\bconflocation#1{#1} \fi

\bibitem[\protect\citeauthoryear{{Abramenko}
  \textit{et~al.}}{2010}]{2010ApJ...712..813A}
\begin{barticle}
\bauthor{\bsnm{{Abramenko}}, \binits{V.}},
\bauthor{\bsnm{{Yurchyshyn}}, \binits{V.}},
\bauthor{\bsnm{{Linker}}, \binits{J.}},
\bauthor{\bsnm{{Miki{\'c}}}, \binits{Z.}},
\bauthor{\bsnm{{Luhmann}}, \binits{J.}},
\bauthor{\bsnm{{Lee}}, \binits{C.O.}}:
\byear{2010},
\batitle{{Low-Latitude Coronal Holes at the Minimum of the 23rd Solar Cycle}}.
\bjtitle{\apj}
\bvolume{712},
\bfpage{813}.
\doiurl{10.1088/0004-637X/712/2/813}.
\adsurl{2010ApJ...712..813A}.
\end{barticle}
\endbibitem

\bibitem[\protect\citeauthoryear{{Antia}
  \textit{et~al.}}{2001}]{2001MNRAS.327.1029A}
\begin{barticle}
\bauthor{\bsnm{{Antia}}, \binits{H.M.}},
\bauthor{\bsnm{{Basu}}, \binits{S.}},
\bauthor{\bsnm{{Hill}}, \binits{F.}},
\bauthor{\bsnm{{Howe}}, \binits{R.}},
\bauthor{\bsnm{{Komm}}, \binits{R.W.}},
\bauthor{\bsnm{{Schou}}, \binits{J.}}:
\byear{2001},
\batitle{{Solar-cycle variation of the sound-speed asphericity from GONG and
  MDI data 1995-2000}}.
\bjtitle{\mnras}
\bvolume{327},
\bfpage{1029}.
\doiurl{10.1046/j.1365-8711.2001.04819.x}.
\adsurl{2001MNRAS.327.1029A}.
\end{barticle}
\endbibitem

\bibitem[\protect\citeauthoryear{{Basu} and
  {Antia}}{2010}]{2010ApJ...717..488B}
\begin{barticle}
\bauthor{\bsnm{{Basu}}, \binits{S.}},
\bauthor{\bsnm{{Antia}}, \binits{H.M.}}:
\byear{2010},
\batitle{{Characteristics of Solar Meridional Flows during Solar Cycle 23}}.
\bjtitle{\apj}
\bvolume{717},
\bfpage{488}.
\doiurl{10.1088/0004-637X/717/1/488}.
\adsurl{2010ApJ...717..488B}.
\end{barticle}
\endbibitem

\bibitem[\protect\citeauthoryear{{Basu}
  \textit{et~al.}}{2012}]{2012ApJ...758...43B}
\begin{barticle}
\bauthor{\bsnm{{Basu}}, \binits{S.}},
\bauthor{\bsnm{{Broomhall}}, \binits{A.-M.}},
\bauthor{\bsnm{{Chaplin}}, \binits{W.J.}},
\bauthor{\bsnm{{Elsworth}}, \binits{Y.}}:
\byear{2012},
\batitle{{Thinning of the Sun's Magnetic Layer: The Peculiar Solar Minimum
  Could Have Been Predicted}}.
\bjtitle{\apj}
\bvolume{758},
\bfpage{43}.
\doiurl{10.1088/0004-637X/758/1/43}.
\adsurl{2012ApJ...758...43B}.
\end{barticle}
\endbibitem

\bibitem[\protect\citeauthoryear{{Bazilevskaya}
  \textit{et~al.}}{2013}]{2013CosRe..51...29B}
\begin{barticle}
\bauthor{\bsnm{{Bazilevskaya}}, \binits{G.A.}},
\bauthor{\bsnm{{Krainev}}, \binits{M.B.}},
\bauthor{\bsnm{{Svirzhevskaya}}, \binits{A.K.}},
\bauthor{\bsnm{{Svirzhevsky}}, \binits{N.S.}}:
\byear{2013},
\batitle{{Galactic cosmic rays and parameters of the interplanetary medium near
  solar activity minima}}.
\bjtitle{Cosmic Research}
\bvolume{51},
\bfpage{29}.
\doiurl{10.1134/S0010952513010012}.
\adsurl{2013CosRe..51...29B}.
\end{barticle}
\endbibitem

\bibitem[\protect\citeauthoryear{{Bazilevskaya}
  \textit{et~al.}}{2014}]{2014SSRv..186..359B}
\begin{barticle}
\bauthor{\bsnm{{Bazilevskaya}}, \binits{G.}},
\bauthor{\bsnm{{Broomhall}}, \binits{A.-M.}},
\bauthor{\bsnm{{Elsworth}}, \binits{Y.}},
\bauthor{\bsnm{{Nakariakov}}, \binits{V.M.}}:
\byear{2014},
\batitle{{A Combined Analysis of the Observational Aspects of the
  Quasi-biennial Oscillation in Solar Magnetic Activity}}.
\bjtitle{\ssr}
\bvolume{186},
\bfpage{359}.
\doiurl{10.1007/s11214-014-0068-0}.
\adsurl{2014SSRv..186..359B}.
\end{barticle}
\endbibitem

\bibitem[\protect\citeauthoryear{{Broomhall}
  \textit{et~al.}}{2009}]{2009ApJ...700L.162B}
\begin{barticle}
\bauthor{\bsnm{{Broomhall}}, \binits{A.-M.}},
\bauthor{\bsnm{{Chaplin}}, \binits{W.J.}},
\bauthor{\bsnm{{Elsworth}}, \binits{Y.}},
\bauthor{\bsnm{{Fletcher}}, \binits{S.T.}},
\bauthor{\bsnm{{New}}, \binits{R.}}:
\byear{2009},
\batitle{{Is the Current Lack of Solar Activity Only Skin Deep?}}
\bjtitle{\apjl}
\bvolume{700},
\bfpage{L162}.
\doiurl{10.1088/0004-637X/700/2/L162}.
\adsurl{2009ApJ...700L.162B}.
\end{barticle}
\endbibitem

\bibitem[\protect\citeauthoryear{{Chaplin}
  \textit{et~al.}}{1996}]{1996SoPh..168....1C}
\begin{barticle}
\bauthor{\bsnm{{Chaplin}}, \binits{W.J.}},
\bauthor{\bsnm{{Elsworth}}, \binits{Y.}},
\bauthor{\bsnm{{Howe}}, \binits{R.}},
\bauthor{\bsnm{{Isaak}}, \binits{G.R.}},
\bauthor{\bsnm{{McLeod}}, \binits{C.P.}},
\bauthor{\bsnm{{Miller}}, \binits{B.A.}},
\bauthor{\bsnm{{van der Raay}}, \binits{H.B.}},
\bauthor{\bsnm{{Wheeler}}, \binits{S.J.}},
\bauthor{\bsnm{{New}}, \binits{R.}}:
\byear{1996},
\batitle{{BiSON Performance}}.
\bjtitle{\solphys}
\bvolume{168},
\bfpage{1}.
\doiurl{10.1007/BF00145821}.
\adsurl{1996SoPh..168....1C}.
\end{barticle}
\endbibitem

\bibitem[\protect\citeauthoryear{{Chaplin}
  \textit{et~al.}}{2004}]{2004MNRAS.352.1102C}
\begin{barticle}
\bauthor{\bsnm{{Chaplin}}, \binits{W.J.}},
\bauthor{\bsnm{{Elsworth}}, \binits{Y.}},
\bauthor{\bsnm{{Isaak}}, \binits{G.R.}},
\bauthor{\bsnm{{Miller}}, \binits{B.A.}},
\bauthor{\bsnm{{New}}, \binits{R.}}:
\byear{2004},
\batitle{{The solar cycle as seen by low-l p-mode frequencies: comparison with
  global and decomposed activity proxies}}.
\bjtitle{\mnras}
\bvolume{352},
\bfpage{1102}.
\doiurl{10.1111/j.1365-2966.2004.07998.x}.
\adsurl{2004MNRAS.352.1102C}.
\end{barticle}
\endbibitem

\bibitem[\protect\citeauthoryear{{Chaplin}
  \textit{et~al.}}{2007}]{2007ApJ...659.1749C}
\begin{barticle}
\bauthor{\bsnm{{Chaplin}}, \binits{W.J.}},
\bauthor{\bsnm{{Elsworth}}, \binits{Y.}},
\bauthor{\bsnm{{Miller}}, \binits{B.A.}},
\bauthor{\bsnm{{Verner}}, \binits{G.A.}},
\bauthor{\bsnm{{New}}, \binits{R.}}:
\byear{2007},
\batitle{{Solar p-Mode Frequencies over Three Solar Cycles}}.
\bjtitle{\apj}
\bvolume{659},
\bfpage{1749}.
\doiurl{10.1086/512543}.
\adsurl{2007ApJ...659.1749C}.
\end{barticle}
\endbibitem

\bibitem[\protect\citeauthoryear{{Clette} and
  {Lef{\`e}vre}}{2012}]{2012JSWSC...2A..06C}
\begin{barticle}
\bauthor{\bsnm{{Clette}}, \binits{F.}},
\bauthor{\bsnm{{Lef{\`e}vre}}, \binits{L.}}:
\byear{2012},
\batitle{{Are the sunspots really vanishing?. Anomalies in solar cycle 23 and
  implications for long-term models and proxies}}.
\bjtitle{J. Space Weather and Space Climate}
\bvolume{2}(\bissue{27}),
\bfpage{A6}.
\doiurl{10.1051/swsc/2012007}.
\adsurl{2012JSWSC...2A..06C}.
\end{barticle}
\endbibitem

\bibitem[\protect\citeauthoryear{{Clette}
  \textit{et~al.}}{2014}]{2014SSRv..186...35C}
\begin{barticle}
\bauthor{\bsnm{{Clette}}, \binits{F.}},
\bauthor{\bsnm{{Svalgaard}}, \binits{L.}},
\bauthor{\bsnm{{Vaquero}}, \binits{J.M.}},
\bauthor{\bsnm{{Cliver}}, \binits{E.W.}}:
\byear{2014},
\batitle{{Revisiting the Sunspot Number. A 400-Year Perspective on the Solar
  Cycle}}.
\bjtitle{\ssr}
\bvolume{186},
\bfpage{35}.
\doiurl{10.1007/s11214-014-0074-2}.
\adsurl{2014SSRv..186...35C}.
\end{barticle}
\endbibitem

\bibitem[\protect\citeauthoryear{{Davies}
  \textit{et~al.}}{2014}]{2014MNRAS.441.3009D}
\begin{barticle}
\bauthor{\bsnm{{Davies}}, \binits{G.R.}},
\bauthor{\bsnm{{Chaplin}}, \binits{W.J.}},
\bauthor{\bsnm{{Elsworth}}, \binits{Y.}},
\bauthor{\bsnm{{Hale}}, \binits{S.J.}}:
\byear{2014},
\batitle{{BiSON data preparation: a correction for differential extinction and
  the weighted averaging of contemporaneous data}}.
\bjtitle{\mnras}
\bvolume{441},
\bfpage{3009}.
\doiurl{10.1093/mnras/stu803}.
\adsurl{2014MNRAS.441.3009D}.
\end{barticle}
\endbibitem

\bibitem[\protect\citeauthoryear{{Dziembowski} and
  {Goode}}{2005}]{2005ApJ...625..548D}
\begin{barticle}
\bauthor{\bsnm{{Dziembowski}}, \binits{W.A.}},
\bauthor{\bsnm{{Goode}}, \binits{P.R.}}:
\byear{2005},
\batitle{{Sources of Oscillation Frequency Increase with Rising Solar
  Activity}}.
\bjtitle{\apj}
\bvolume{625},
\bfpage{548}.
\doiurl{10.1086/429712}.
\adsurl{2005ApJ...625..548D}.
\end{barticle}
\endbibitem

\bibitem[\protect\citeauthoryear{{Elsworth}
  \textit{et~al.}}{1990}]{1990Natur.345..322E}
\begin{barticle}
\bauthor{\bsnm{{Elsworth}}, \binits{Y.}},
\bauthor{\bsnm{{Howe}}, \binits{R.}},
\bauthor{\bsnm{{Isaak}}, \binits{G.R.}},
\bauthor{\bsnm{{McLeod}}, \binits{C.P.}},
\bauthor{\bsnm{{New}}, \binits{R.}}:
\byear{1990},
\batitle{{Variation of low-order acoustic solar oscillations over the solar
  cycle}}.
\bjtitle{\nat}
\bvolume{345},
\bfpage{322}.
\doiurl{10.1038/345322a0}.
\adsurl{http://cdsads.u-strasbg.fr/abs/1990Natur.345..322E}.
\end{barticle}
\endbibitem

\bibitem[\protect\citeauthoryear{{Fletcher}
  \textit{et~al.}}{2009}]{2009ApJ...694..144F}
\begin{barticle}
\bauthor{\bsnm{{Fletcher}}, \binits{S.T.}},
\bauthor{\bsnm{{Chaplin}}, \binits{W.J.}},
\bauthor{\bsnm{{Elsworth}}, \binits{Y.}},
\bauthor{\bsnm{{New}}, \binits{R.}}:
\byear{2009},
\batitle{{Efficient Pseudo-Global Fitting for Helioseismic Data}}.
\bjtitle{\apj}
\bvolume{694},
\bfpage{144}.
\doiurl{10.1088/0004-637X/694/1/144}.
\adsurl{2009ApJ...694..144F}.
\end{barticle}
\endbibitem

\bibitem[\protect\citeauthoryear{{Hathaway}}{2010}]{2010LRSP....7....1H}
\begin{barticle}
\bauthor{\bsnm{{Hathaway}}, \binits{D.H.}}:
\byear{2010},
\batitle{{The Solar Cycle}}.
\bjtitle{Living Rev. Solar Physics}
\bvolume{7},
\bfpage{1}.
\doiurl{10.12942/lrsp-2010-1}.
\adsurl{2010LRSP....7....1H}.
\end{barticle}
\endbibitem

\bibitem[\protect\citeauthoryear{{Hathaway} and
  {Rightmire}}{2010}]{2010Sci...327.1350H}
\begin{barticle}
\bauthor{\bsnm{{Hathaway}}, \binits{D.H.}},
\bauthor{\bsnm{{Rightmire}}, \binits{L.}}:
\byear{2010},
\batitle{{Variations in the Sun's Meridional Flow over a Solar Cycle}}.
\bjtitle{Science}
\bvolume{327}.
\doiurl{10.1126/science.1181990}.
\adsurl{2010Sci...327.1350H}.
\end{barticle}
\endbibitem

\bibitem[\protect\citeauthoryear{{Howe}, {Komm}, and
  {Hill}}{1999}]{1999ApJ...524.1084H}
\begin{barticle}
\bauthor{\bsnm{{Howe}}, \binits{R.}},
\bauthor{\bsnm{{Komm}}, \binits{R.}},
\bauthor{\bsnm{{Hill}}, \binits{F.}}:
\byear{1999},
\batitle{{Solar Cycle Changes in GONG P-Mode Frequencies, 1995-1998}}.
\bjtitle{\apj}
\bvolume{524},
\bfpage{1084}.
\doiurl{10.1086/307851}.
\adsurl{1999ApJ...524.1084H}.
\end{barticle}
\endbibitem

\bibitem[\protect\citeauthoryear{{Howe}, {Komm}, and
  {Hill}}{2002}]{2002ApJ...580.1172H}
\begin{barticle}
\bauthor{\bsnm{{Howe}}, \binits{R.}},
\bauthor{\bsnm{{Komm}}, \binits{R.W.}},
\bauthor{\bsnm{{Hill}}, \binits{F.}}:
\byear{2002},
\batitle{{Localizing the Solar Cycle Frequency Shifts in Global p-Modes}}.
\bjtitle{\apj}
\bvolume{580},
\bfpage{1172}.
\doiurl{10.1086/343892}.
\adsurl{2002ApJ...580.1172H}.
\end{barticle}
\endbibitem

\bibitem[\protect\citeauthoryear{{Howe}
  \textit{et~al.}}{2013}]{2013ApJ...767L..20H}
\begin{barticle}
\bauthor{\bsnm{{Howe}}, \binits{R.}},
\bauthor{\bsnm{{Christensen-Dalsgaard}}, \binits{J.}},
\bauthor{\bsnm{{Hill}}, \binits{F.}},
\bauthor{\bsnm{{Komm}}, \binits{R.}},
\bauthor{\bsnm{{Larson}}, \binits{T.P.}},
\bauthor{\bsnm{{Rempel}}, \binits{M.}},
\bauthor{\bsnm{{Schou}}, \binits{J.}},
\bauthor{\bsnm{{Thompson}}, \binits{M.J.}}:
\byear{2013},
\batitle{{The High-latitude Branch of the Solar Torsional Oscillation in the
  Rising Phase of Cycle 24}}.
\bjtitle{\apjl}
\bvolume{767},
\bfpage{L20}.
\doiurl{10.1088/2041-8205/767/1/L20}.
\adsurl{2013ApJ...767L..20H}.
\end{barticle}
\endbibitem

\bibitem[\protect\citeauthoryear{{Jain}, {Tripathy}, and
  {Hill}}{2009}]{2009ApJ...695.1567J}
\begin{barticle}
\bauthor{\bsnm{{Jain}}, \binits{K.}},
\bauthor{\bsnm{{Tripathy}}, \binits{S.C.}},
\bauthor{\bsnm{{Hill}}, \binits{F.}}:
\byear{2009},
\batitle{{Solar Activity Phases and Intermediate-Degree Mode Frequencies}}.
\bjtitle{\apj}
\bvolume{695},
\bfpage{1567}.
\doiurl{10.1088/0004-637X/695/2/1567}.
\adsurl{2009ApJ...695.1567J}.
\end{barticle}
\endbibitem

\bibitem[\protect\citeauthoryear{{Jain} and
  {Roberts}}{1994}]{1994SoPh..152..261J}
\begin{barticle}
\bauthor{\bsnm{{Jain}}, \binits{R.}},
\bauthor{\bsnm{{Roberts}}, \binits{B.}}:
\byear{1994},
\batitle{{Solar cycle variations in p-modes and chromospheric magnetism}}.
\bjtitle{\solphys}
\bvolume{152},
\bfpage{261}.
\doiurl{10.1007/BF01473213}.
\adsurl{1994SoPh..152..261J}.
\end{barticle}
\endbibitem

\bibitem[\protect\citeauthoryear{{Jain}
  \textit{et~al.}}{2012}]{2012A&A...545A..73J}
\begin{barticle}
\bauthor{\bsnm{{Jain}}, \binits{R.}},
\bauthor{\bsnm{{Tripathy}}, \binits{S.C.}},
\bauthor{\bsnm{{Watson}}, \binits{F.T.}},
\bauthor{\bsnm{{Fletcher}}, \binits{L.}},
\bauthor{\bsnm{{Jain}}, \binits{K.}},
\bauthor{\bsnm{{Hill}}, \binits{F.}}:
\byear{2012},
\batitle{{Variation of solar oscillation frequencies in solar cycle 23 and
  their relation to sunspot area and number}}.
\bjtitle{\aap}
\bvolume{545},
\bfpage{A73}.
\doiurl{10.1051/0004-6361/201219876}.
\adsurl{2012A\%26A...545A..73J}.
\end{barticle}
\endbibitem

\bibitem[\protect\citeauthoryear{{Javaraiah}}{2011}]{2011SoPh..270..463J}
\begin{barticle}
\bauthor{\bsnm{{Javaraiah}}, \binits{J.}}:
\byear{2011},
\batitle{{Long-Term Variations in the Growth and Decay Rates of Sunspot
  Groups}}.
\bjtitle{\solphys}
\bvolume{270},
\bfpage{463}.
\doiurl{10.1007/s11207-011-9768-8}.
\adsurl{2011SoPh..270..463J}.
\end{barticle}
\endbibitem

\bibitem[\protect\citeauthoryear{{Jimenez-Reyes}
  \textit{et~al.}}{1998}]{1998A&A...329.1119J}
\begin{barticle}
\bauthor{\bsnm{{Jimenez-Reyes}}, \binits{S.J.}},
\bauthor{\bsnm{{Regulo}}, \binits{C.}},
\bauthor{\bsnm{{Palle}}, \binits{P.L.}},
\bauthor{\bsnm{{Roca Cortes}}, \binits{T.}}:
\byear{1998},
\batitle{{Solar activity cycle frequency shifts of low-degree p-modes}}.
\bjtitle{\aap}
\bvolume{329},
\bfpage{1119}.
\adsurl{1998A\%26A...329.1119J}.
\end{barticle}
\endbibitem

\bibitem[\protect\citeauthoryear{{Kilcik}
  \textit{et~al.}}{2011}]{2011ApJ...731...30K}
\begin{barticle}
\bauthor{\bsnm{{Kilcik}}, \binits{A.}},
\bauthor{\bsnm{{Yurchyshyn}}, \binits{V.B.}},
\bauthor{\bsnm{{Abramenko}}, \binits{V.}},
\bauthor{\bsnm{{Goode}}, \binits{P.R.}},
\bauthor{\bsnm{{Ozguc}}, \binits{A.}},
\bauthor{\bsnm{{Rozelot}}, \binits{J.P.}},
\bauthor{\bsnm{{Cao}}, \binits{W.}}:
\byear{2011},
\batitle{{Time Distributions of Large and Small Sunspot Groups Over Four Solar
  Cycles}}.
\bjtitle{\apj}
\bvolume{731},
\bfpage{30}.
\doiurl{10.1088/0004-637X/731/1/30}.
\adsurl{2011ApJ...731...30K}.
\end{barticle}
\endbibitem

\bibitem[\protect\citeauthoryear{{King} and
  {Papitashvili}}{2005}]{2005JGRA..110.2104K}
\begin{barticle}
\bauthor{\bsnm{{King}}, \binits{J.H.}},
\bauthor{\bsnm{{Papitashvili}}, \binits{N.E.}}:
\byear{2005},
\batitle{{Solar wind spatial scales in and comparisons of hourly Wind and ACE
  plasma and magnetic field data}}.
\bjtitle{J. Geophys. Res. (Space Physics)}
\bvolume{110},
\bfpage{2104}.
\doiurl{10.1029/2004JA010649}.
\adsurl{2005JGRA..110.2104K}.
\end{barticle}
\endbibitem

\bibitem[\protect\citeauthoryear{{Lef{\`e}vre} and
  {Clette}}{2011}]{2011A&A...536L..11L}
\begin{barticle}
\bauthor{\bsnm{{Lef{\`e}vre}}, \binits{L.}},
\bauthor{\bsnm{{Clette}}, \binits{F.}}:
\byear{2011},
\batitle{{A global small sunspot deficit at the base of the index anomalies of
  solar cycle 23}}.
\bjtitle{\aap}
\bvolume{536},
\bfpage{L11}.
\doiurl{10.1051/0004-6361/201118034}.
\adsurl{2011A\%26A...536L..11L}.
\end{barticle}
\endbibitem

\bibitem[\protect\citeauthoryear{{Libbrecht} and
  {Woodard}}{1990}]{1990Natur.345..779L}
\begin{barticle}
\bauthor{\bsnm{{Libbrecht}}, \binits{K.G.}},
\bauthor{\bsnm{{Woodard}}, \binits{M.F.}}:
\byear{1990},
\batitle{{Solar-cycle effects on solar oscillation frequencies}}.
\bjtitle{\nat}
\bvolume{345},
\bfpage{779}.
\doiurl{10.1038/345779a0}.
\adsurl{1990Natur.345..779L}.
\end{barticle}
\endbibitem

\bibitem[\protect\citeauthoryear{{Livingston}}{2002}]{2002SoPh..207...41L}
\begin{barticle}
\bauthor{\bsnm{{Livingston}}, \binits{W.}}:
\byear{2002},
\batitle{{Sunspots Observed to Physically Weaken in 2000-2001}}.
\bjtitle{\solphys}
\bvolume{207},
\bfpage{41}.
\doiurl{10.1023/A:1015555000456}.
\adsurl{2002SoPh..207...41L}.
\end{barticle}
\endbibitem

\bibitem[\protect\citeauthoryear{{Livingston}, {Penn}, and
  {Svalgaard}}{2012}]{2012ApJ...757L...8L}
\begin{barticle}
\bauthor{\bsnm{{Livingston}}, \binits{W.}},
\bauthor{\bsnm{{Penn}}, \binits{M.J.}},
\bauthor{\bsnm{{Svalgaard}}, \binits{L.}}:
\byear{2012},
\batitle{{Decreasing Sunspot Magnetic Fields Explain Unique 10.7 cm Radio
  Flux}}.
\bjtitle{\apjl}
\bvolume{757},
\bfpage{L8}.
\doiurl{10.1088/2041-8205/757/1/L8}.
\adsurl{2012ApJ...757L...8L}.
\end{barticle}
\endbibitem

\bibitem[\protect\citeauthoryear{{Mavromichalaki}
  \textit{et~al.}}{2005}]{2005AdSpR..35..410M}
\begin{barticle}
\bauthor{\bsnm{{Mavromichalaki}}, \binits{H.}},
\bauthor{\bsnm{{Petropoulos}}, \binits{B.}},
\bauthor{\bsnm{{Plainaki}}, \binits{C.}},
\bauthor{\bsnm{{Dionatos}}, \binits{O.}},
\bauthor{\bsnm{{Zouganelis}}, \binits{I.}}:
\byear{2005},
\batitle{{Coronal index as a solar activity index applied to space weather}}.
\bjtitle{Adv. Space Res.}
\bvolume{35},
\bfpage{410}.
\doiurl{10.1016/j.asr.2005.01.084}.
\adsurl{2005AdSpR..35..410M}.
\end{barticle}
\endbibitem

\bibitem[\protect\citeauthoryear{{Moreno-Insertis} and
  {Solanki}}{2000}]{2000MNRAS.313..411M}
\begin{barticle}
\bauthor{\bsnm{{Moreno-Insertis}}, \binits{F.}},
\bauthor{\bsnm{{Solanki}}, \binits{S.K.}}:
\byear{2000},
\batitle{{Distribution of magnetic flux on the solar surface and low-degree
  p-modes}}.
\bjtitle{\mnras}
\bvolume{313},
\bfpage{411}.
\doiurl{10.1046/j.1365-8711.2000.03246.x}.
\adsurl{2000MNRAS.313..411M}.
\end{barticle}
\endbibitem

\bibitem[\protect\citeauthoryear{{Nagovitsyn}, {Pevtsov}, and
  {Livingston}}{2012}]{2012ApJ...758L..20N}
\begin{barticle}
\bauthor{\bsnm{{Nagovitsyn}}, \binits{Y.A.}},
\bauthor{\bsnm{{Pevtsov}}, \binits{A.A.}},
\bauthor{\bsnm{{Livingston}}, \binits{W.C.}}:
\byear{2012},
\batitle{{On a Possible Explanation of the Long-term Decrease in Sunspot Field
  Strength}}.
\bjtitle{\apjl}
\bvolume{758},
\bfpage{L20}.
\doiurl{10.1088/2041-8205/758/1/L20}.
\adsurl{2012ApJ...758L..20N}.
\end{barticle}
\endbibitem

\bibitem[\protect\citeauthoryear{{Nigam} and
  {Kosovichev}}{1998}]{1998ApJ...505L..51N}
\begin{barticle}
\bauthor{\bsnm{{Nigam}}, \binits{R.}},
\bauthor{\bsnm{{Kosovichev}}, \binits{A.G.}}:
\byear{1998},
\batitle{{Measuring the Sun's Eigenfrequencies from Velocity and Intensity
  Helioseismic Spectra: Asymmetrical Line Profile-fitting Formula}}.
\bjtitle{\apjl}
\bvolume{505},
\bfpage{L51}.
\doiurl{10.1086/311594}.
\adsurl{1998ApJ...505L..51N}.
\end{barticle}
\endbibitem

\bibitem[\protect\citeauthoryear{{Palle}, {Regulo}, and {Roca
  Cortes}}{1989}]{1989A&A...224..253P}
\begin{barticle}
\bauthor{\bsnm{{Palle}}, \binits{P.L.}},
\bauthor{\bsnm{{Regulo}}, \binits{C.}},
\bauthor{\bsnm{{Roca Cortes}}, \binits{T.}}:
\byear{1989},
\batitle{{Solar cycle induced variations of the low L solar acoustic
  spectrum}}.
\bjtitle{\aap}
\bvolume{224},
\bfpage{253}.
\adsurl{1989A\%26A...224..253P}.
\end{barticle}
\endbibitem

\bibitem[\protect\citeauthoryear{{Penn} and
  {Livingston}}{2006}]{2006ApJ...649L..45P}
\begin{barticle}
\bauthor{\bsnm{{Penn}}, \binits{M.J.}},
\bauthor{\bsnm{{Livingston}}, \binits{W.}}:
\byear{2006},
\batitle{{Temporal Changes in Sunspot Umbral Magnetic Fields and
  Temperatures}}.
\bjtitle{\apjl}
\bvolume{649},
\bfpage{L45}.
\doiurl{10.1086/508345}.
\adsurl{2006ApJ...649L..45P}.
\end{barticle}
\endbibitem

\bibitem[\protect\citeauthoryear{{Penn} and
  {Livingston}}{2011}]{2011IAUS..273..126P}
\begin{bchapter}
\bauthor{\bsnm{{Penn}}, \binits{M.J.}},
\bauthor{\bsnm{{Livingston}}, \binits{W.}}:
\byear{2011},
\bctitle{{Long-term evolution of sunspot magnetic fields}}.
In: \beditor{\bsnm{{Prasad Choudhary}}, \binits{D.}},
\beditor{\bsnm{{Strassmeier}}, \binits{K.G.}} (eds.)
\bbtitle{IAU Symposium},
\bsertitle{IAU Symposium}
\bseriesno{273},
\bfpage{126}.
\doiurl{10.1017/S1743921311015122}.
\adsurl{2011IAUS..273..126P}.
\end{bchapter}
\endbibitem

\bibitem[\protect\citeauthoryear{{Pevtsov}
  \textit{et~al.}}{2011}]{2011ApJ...742L..36P}
\begin{barticle}
\bauthor{\bsnm{{Pevtsov}}, \binits{A.A.}},
\bauthor{\bsnm{{Nagovitsyn}}, \binits{Y.A.}},
\bauthor{\bsnm{{Tlatov}}, \binits{A.G.}},
\bauthor{\bsnm{{Rybak}}, \binits{A.L.}}:
\byear{2011},
\batitle{{Long-term Trends in Sunspot Magnetic Fields}}.
\bjtitle{\apjl}
\bvolume{742},
\bfpage{L36}.
\doiurl{10.1088/2041-8205/742/2/L36}.
\adsurl{2011ApJ...742L..36P}.
\end{barticle}
\endbibitem

\bibitem[\protect\citeauthoryear{{Pevtsov}
  \textit{et~al.}}{2014}]{2014SoPh..289..593P}
\begin{barticle}
\bauthor{\bsnm{{Pevtsov}}, \binits{A.A.}},
\bauthor{\bsnm{{Bertello}}, \binits{L.}},
\bauthor{\bsnm{{Tlatov}}, \binits{A.G.}},
\bauthor{\bsnm{{Kilcik}}, \binits{A.}},
\bauthor{\bsnm{{Nagovitsyn}}, \binits{Y.A.}},
\bauthor{\bsnm{{Cliver}}, \binits{E.W.}}:
\byear{2014},
\batitle{{Cyclic and Long-Term Variation of Sunspot Magnetic Fields}}.
\bjtitle{\solphys}
\bvolume{289},
\bfpage{593}.
\doiurl{10.1007/s11207-012-0220-5}.
\adsurl{2014SoPh..289..593P}.
\end{barticle}
\endbibitem

\bibitem[\protect\citeauthoryear{{Rezaei}, {Beck}, and
  {Schmidt}}{2012}]{2012A&A...541A..60R}
\begin{barticle}
\bauthor{\bsnm{{Rezaei}}, \binits{R.}},
\bauthor{\bsnm{{Beck}}, \binits{C.}},
\bauthor{\bsnm{{Schmidt}}, \binits{W.}}:
\byear{2012},
\batitle{{Variation in sunspot properties between 1999 and 2011 as observed
  with the Tenerife Infrared Polarimeter}}.
\bjtitle{\aap}
\bvolume{541},
\bfpage{A60}.
\doiurl{10.1051/0004-6361/201118635}.
\adsurl{2012A\%26A...541A..60R}.
\end{barticle}
\endbibitem

\bibitem[\protect\citeauthoryear{{Roberts} and
  {Campbell}}{1986}]{1986Natur.323..603R}
\begin{barticle}
\bauthor{\bsnm{{Roberts}}, \binits{B.}},
\bauthor{\bsnm{{Campbell}}, \binits{W.R.}}:
\byear{1986},
\batitle{{Magnetic field corrections to solar oscillation frequencies}}.
\bjtitle{\nat}
\bvolume{323},
\bfpage{603}.
\doiurl{10.1038/323603a0}.
\adsurl{1986Natur.323..603R}.
\end{barticle}
\endbibitem

\bibitem[\protect\citeauthoryear{{Salabert}, {Garcia}, and
  {Turck-Chieze}}{2015}]{2015arXiv150207607S}
\begin{botherref}
\oauthor{\bsnm{{Salabert}}, \binits{D.}},
\oauthor{\bsnm{{Garcia}}, \binits{R.A.}},
\oauthor{\bsnm{{Turck-Chieze}}, \binits{S.}}:
2015,
{Seismic sensitivity to sub-surface solar activity from 18 years of GOLF/SoHO
  observations}.
\textit{ArXiv e-prints}.
\adsurl{2015arXiv150207607S}.
\end{botherref}
\endbibitem

\bibitem[\protect\citeauthoryear{{Tapping}}{1987}]{1987JGR....92..829T}
\begin{barticle}
\bauthor{\bsnm{{Tapping}}, \binits{K.F.}}:
\byear{1987},
\batitle{{Recent solar radio astronomy at centimeter wavelengths - The temporal
  variability of the 10.7-cm flux}}.
\bjtitle{\jgr}
\bvolume{92},
\bfpage{829}.
\doiurl{10.1029/JD092iD01p00829}.
\adsurl{1987JGR....92..829T}.
\end{barticle}
\endbibitem

\bibitem[\protect\citeauthoryear{{Tlatov}}{2013}]{2013Ge&Ae..53..953T}
\begin{barticle}
\bauthor{\bsnm{{Tlatov}}, \binits{A.G.}}:
\byear{2013},
\batitle{{Long-term variations in sunspot characteristics}}.
\bjtitle{Geomagnetism and Aeronomy}
\bvolume{53},
\bfpage{953}.
\doiurl{10.1134/S0016793213080264}.
\adsurl{2013Ge\%26Ae..53..953T}.
\end{barticle}
\endbibitem

\bibitem[\protect\citeauthoryear{{Tripathy}
  \textit{et~al.}}{2000}]{2000JApA...21..357T}
\begin{barticle}
\bauthor{\bsnm{{Tripathy}}, \binits{S.C.}},
\bauthor{\bsnm{{Kumar}}, \binits{B.}},
\bauthor{\bsnm{{Jain}}, \binits{K.}},
\bauthor{\bsnm{{Bhatnagar}}, \binits{A.}}:
\byear{2000},
\batitle{{Observation of Hysteresis Between Solar Activity Indicators and
  p-mode Frequency Shifts for Solar Cycle 22}}.
\bjtitle{J. Astrophys. Astron.}
\bvolume{21},
\bfpage{357}.
\doiurl{10.1007/BF02702424}.
\adsurl{2000JApA...21..357T}.
\end{barticle}
\endbibitem

\bibitem[\protect\citeauthoryear{{Tripathy}
  \textit{et~al.}}{2001}]{2001SoPh..200....3T}
\begin{barticle}
\bauthor{\bsnm{{Tripathy}}, \binits{S.C.}},
\bauthor{\bsnm{{Kumar}}, \binits{B.}},
\bauthor{\bsnm{{Jain}}, \binits{K.}},
\bauthor{\bsnm{{Bhatnagar}}, \binits{A.}}:
\byear{2001},
\batitle{{Analysis of hysteresis effect in p-mode frequency shifts and solar
  activity indices}}.
\bjtitle{\solphys}
\bvolume{200},
\bfpage{3}.
\doiurl{10.1023/A:1010318428454}.
\adsurl{2001SoPh..200....3T}.
\end{barticle}
\endbibitem

\bibitem[\protect\citeauthoryear{{Watson}, {Fletcher}, and
  {Marshall}}{2011}]{2011A&A...533A..14W}
\begin{barticle}
\bauthor{\bsnm{{Watson}}, \binits{F.T.}},
\bauthor{\bsnm{{Fletcher}}, \binits{L.}},
\bauthor{\bsnm{{Marshall}}, \binits{S.}}:
\byear{2011},
\batitle{{Evolution of sunspot properties during solar cycle 23}}.
\bjtitle{\aap}
\bvolume{533},
\bfpage{A14}.
\doiurl{10.1051/0004-6361/201116655}.
\adsurl{2011A\%26A...533A..14W}.
\end{barticle}
\endbibitem

\bibitem[\protect\citeauthoryear{{Watson}, {Penn}, and
  {Livingston}}{2014}]{2014ApJ...787...22W}
\begin{barticle}
\bauthor{\bsnm{{Watson}}, \binits{F.T.}},
\bauthor{\bsnm{{Penn}}, \binits{M.J.}},
\bauthor{\bsnm{{Livingston}}, \binits{W.}}:
\byear{2014},
\batitle{{A Multi-instrument Analysis of Sunspot Umbrae}}.
\bjtitle{\apj}
\bvolume{787},
\bfpage{22}.
\doiurl{10.1088/0004-637X/787/1/22}.
\adsurl{2014ApJ...787...22W}.
\end{barticle}
\endbibitem

\bibitem[\protect\citeauthoryear{{Woodard} and
  {Noyes}}{1985}]{1985Natur.318..449W}
\begin{barticle}
\bauthor{\bsnm{{Woodard}}, \binits{M.F.}},
\bauthor{\bsnm{{Noyes}}, \binits{R.W.}}:
\byear{1985},
\batitle{{Change of Solar Oscillation Eigenfrequencies with the Solar Cycle}}.
\bjtitle{Nat.}
\bvolume{318},
\bfpage{449}.
\adsurl{http://ukads.nottingham.ac.uk/abs/1985Natur.318..449W}.
\end{barticle}
\endbibitem

\end{thebibliography}

\end{article}
\end{document}